\documentclass[reprint,superscriptaddress,amsmath,amssymb,aps,prl,floatfix,longbibliography]{revtex4-1}

\usepackage{amsmath,amssymb,graphicx,mathrsfs,amsfonts}
\usepackage{floatrow, placeins}

\newcommand{\bra}[1]{\ensuremath{\left\langle {#1} \right|}}
\newcommand{\ket}[1]{\ensuremath{\left|  #1 \right\rangle}}

\draft 

\begin{document}

\title{Coupling atoms to cavities with narrow linewidth optical transitions: Applications to frequency metrology} %

\author{Matthew A. Norcia}
\affiliation{JILA, NIST, and Dept. of Physics, University of Colorado, 440 UCB, Boulder, CO  80309, USA}

\date{\today}

\begin{abstract}
{\bf 
\noindent 
Narrow linewidth optical atomic transitions provide a valuable resource for frequency metrology, and form the basis of today's most precise and accurate clocks.  Recent experiments have demonstrated that ensembles of atoms can be interfaced with the mode of an optical cavity using such transitions, and that atom-cavity interactions can dominate over decoherence processes even when the atomic transition that mediates the interactions is very weak.  This scenario enables new opportunities for optical frequency metrology, including techniques for nondestructive readout and entanglement enhancement for optical lattice clocks, methods for cavity-enhanced laser frequency stabilization, and high-precision active optical frequency references based on superradiant emission.  This tutorial provides a pedagogical description of the physics governing atom-cavity coupling with narrow linewidth optical transitions, and describes several examples of applications to optical frequency metrology.  

}
\end{abstract}


\maketitle

\section{Introduction}

A core facet of atomic physics is the study of interactions between atoms and electromagnetic radiation.  While such studies can be performed in free space, the use of a resonator, or cavity, for the radiation can greatly enhance the strength of its interaction with the atoms.  This scenario enables a wealth of applications, including fundamental studies of measurement and decoherence in quantum mechanics \cite{raimond2001manipulating}, quantum communications \cite{kimble2008quantum, kuhn2002deterministic}, studies of optomechanical effects \cite{black2003observation, purdy2010tunable}, and of interacting quantum particles \cite{BGB2010, davis2019photon, norcia2018cavity, chang2018colloquium}. 

The physics of atoms in cavities can be extended to a range of atomic and atom-like systems.  For example, experimental implementations with atom-like systems range from solid-state emitters coupled to nanophotonic cavities \cite{pelton2002efficient}, to superconducting qubits coupled to microwave-domain resonators \cite{wallraff2004strong}.  Cavity systems with actual atoms have been explored in a wide range of parameter regimes, with atoms coupled to resonators operating either in the microwave \cite{raimond2001manipulating} or optical domain \cite{miller2005trapped}.  An emerging direction is the coupling of atoms to cavities using narrow and ultranarrow linewidth optical transitions \cite{dibos2018atomic, PhysRevLett.114.093002, christensen2015nonlinear, braverman2019near, norcia2015strong, Laske2019, norcia2016superradiance}, which will be the subject of this tutorial.  

Typical optical transitions -- those that are not forbidden by dipole selection rules -- have linewidths of a few MHz.  Certain atomic species, such as alkaline earth atoms (which have two valence electrons), also host optical transitions that are nearly forbidden by dipole selection rules and have much narrower decay linewidths ranging from sub-mHz to hundreds of kHz.  These narrow linewidth transitions (as well as similar transitions in ions) have enabled dramatic progress in optical frequency metrology, and currently form the basis of the most precise and accurate atomic clocks \cite{schioppo2017ultrastable, Nicholson2015, origlia2018high, ushijima2015cryogenic, chou2010frequency}.  Combining narrow linewidth optical transitions with the toolkit of optical cavities provides promising new directions for further improvements in optical frequency metrology.  

This tutorial will discuss three such directions: cavity-enhanced laser frequency stabilization techniques based on atoms with narrow linewidth transitions, the use of narrow linewidth transitions and optical cavities for nondestructive readout and spin squeezing in optical clocks,  and active frequency references based on superradiant emission from ultra-narrow atomic transitions.  Before presenting these applications, I will provide a brief introduction to the language and key physical concepts of atoms in cavities, with a focus on the implications of operating with narrow linewidth transitions.

\noindent \subsection{A brief introduction to atoms in cavities}

The following section provides a brief overview of the key physics and parameters that govern a generic atom-cavity system.  Further details and intuition can be found in sources such as \cite{tanji2011interaction, haroche2006exploring, steck2007quantum}.  The focus of this discussion is on important scalings and their implications for using narrow linewidth optical transitions.

A typical goal of coupling atoms to a cavity  is to realize a situation where an atom (or ensemble of atoms) interacts primarily with a single mode of electromagnetic radiation.  The cavity supports discrete resonant modes, each with a well-defined frequency and spatial profile.  Each of these modes is analogous to a harmonic oscillator, with a ladder of evenly spaced energy levels representing integer numbers of photons occupying the mode.  When an atom is located within one of these modes whose resonance frequency is near that of the atomic transition, then the interaction between the atom and this mode may dominate over coupling to other modes of the cavity or the environment.  

\begin{figure}[!htb]
\centering
\includegraphics[width=3.5in, ]{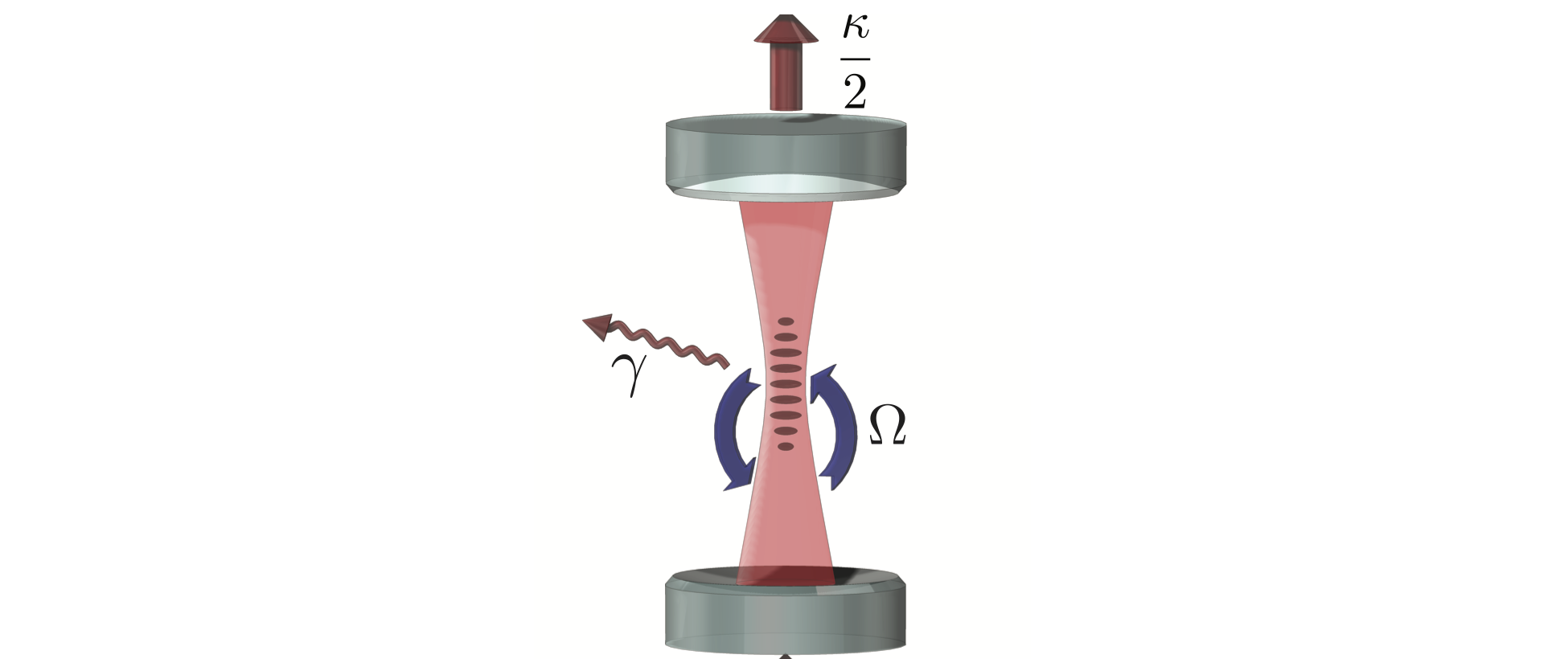}
\caption{An atom-cavity system with $N$ atoms.  Atoms exchange excitations with the cavity mode at a rate $\Omega = 2 g \sqrt{N}$.  Atoms can decay into free-space via single-particle spontaneous emission at a rate $\gamma$, leading to loss of coherence.  Photons in the cavity leak out through the mirrors at rate $\kappa$, providing an efficient channel for collection and transmission of collective information.  }
\label{fig:QEDsystem}
\end{figure}

A single atom interacting with an optical cavity can be described by the so-called Jaynes-Cummings Hamiltonian \cite{jaynes1963comparison}:

$$ \hat{H} = \hbar \omega_a \hat{\sigma}^+ \hat{\sigma}^- + \hbar \omega \hat{c}^\dagger \hat{c} + \hbar g (\hat{\sigma}^- \hat{c}^\dagger + \hat{\sigma}^+ \hat{c})$$

\noindent The first term corresponds to the energy of the atom --- $\omega_a$ is the atomic transition frequency and $\hat{\sigma}^+ =  \ket{e}\bra{g}$ ($\hat{\sigma}^- = \ket{g}\bra{e}$) is the atomic raising (lowering) operator.  The second term represents the energy of the cavity field --- $\omega$ is the cavity resonance frequency, and $\hat{c}^\dagger$ ($\hat{c}$) is the photon creation (annihilation) operators.  Finally, the third term represents the interaction between the atom and the cavity.  The atom can go from $\ket{e}$ to $\ket{g}$ by emitting a photon, or from $\ket{g}$ to $\ket{e}$ by absorbing one.  The frequency at which this happens is determined by $g$, the single particle coupling.   

We can rewrite this Hamiltonian in a rotating frame at the atomic transition frequency $\omega_a$ as: 

$$ \hat{H} = \hbar \delta_c \hat{c}^\dagger \hat{c} + \hbar g (\hat{\sigma}^- \hat{c}^\dagger + \hat{\sigma}^+ \hat{c})$$

\noindent where $\delta_c = \omega-\omega_a$ is the cavity detuning .  

In addition to the coherent exchange of excitations between the atom and cavity, we must consider the coupling of the system to the environment.  The atom can decay from $\ket{e}$ to $\ket{g}$ at a rate $\gamma$ by emitting a photon into a mode other than that of the cavity.  If the cavity subtends a large solid angle, this emission rate can be supressed below its free space value \cite{kleppner1981inhibited}. However, in the optical domain that is the focus here, this suppression is typically a minor effect.  In addition to free-space atomic decay, photons in the cavity decay at a rate $\kappa$, the cavity linewidth, ideally by transmission through the cavity mirrors.  These processes are illustrated in Fig.~\ref{fig:QEDsystem}.   The relative strength of the coherent interactions between the atoms and the cavity to these dissipative processes determines the regime in which an experiment operates.

The single atom coupling $g$ determines the strength of coherent interactions between an atom and the cavity, and can be expressed as:

$$\hbar g = -\hat{\varepsilon} \cdot \vec{d}\sqrt{\frac{\omega}{2 \epsilon_0 \hbar V.}}$$

\noindent Here, $\hat{\varepsilon}$ represents the polarization of the cavity field, $\vec{d}$ is the dipole matrix element of the atomic transition, and $V$ is the volume of the cavity mode (the cross-sectional area at the location of the atoms times the cavity length). For simplicity, I assume here that the cavity mode is uniform over the volume $V$.  Physically, $\hbar g$ represents the energy $\vec{d} \cdot \vec{E}$ associated with the atomic dipole interacting with the field of a single photon in the cavity.

Importantly, $g$ depends only on atomic properties (the dipole matrix element $\vec d$, and the atomic transition frequency $\omega_a \simeq \omega$), and the volume of the cavity mode.  A short cavity with a tightly focused mode will thus have a large $g$, while a long cavity with a spatially broad mode will have a small value of $g$.  Note that the quality of the mirror coatings do not affect the value of $g$. 

For certain applications, many atoms can be placed inside the cavity mode to enhance the rate of coupling.  For $N$ atoms uniformly coupled to the cavity mode with strength $g$, the Hamiltonian becomes 

$$ \hat{H} = \hbar \delta_c \hat{c}^\dagger \hat{c} + \hbar g (\hat{J}_- \hat{c}^\dagger + \hat{J}_+ \hat{c})$$

\noindent where $\hat{J}_\pm = \frac{1}{2}\sum_{i=1}^N \hat{\sigma}^\pm_{i}$ are the collective raising and lowering operators for the atoms.  In applications where the atom-cavity system is subjected to a weak probe, and the number of atomic excitations is much less than the total number of atoms, we can approximate this Hamiltonian \cite{PhysRevA.89.043837, holstein1940field} as

\begin{equation}
\hat{H} = \hbar \delta_c \hat{c}^\dagger \hat{c} + \hbar g \sqrt{N}(\hat{a} \hat{c}^\dagger + \hat{a}^\dagger \hat{c})
\end{equation}

\noindent where we replace $\hat{J}_\pm$ with creation and annihilation operators $\hat{a} = J_+/\sqrt{N}$ and $\hat{a}^\dagger = J_-/\sqrt{N}$.  

When there are many spins inside the cavity, the rate at which a photon is exchanged between the cavity and the atoms is enhanced from $2g$ to $\Omega = 2 g \sqrt{N}$, the collective vacuum Rabi frequency.  This collective enhancement results from the situation where many atoms interact with the same cavity mode, and each experiences the effects of the field radiated by all other atoms.  

Strong coupling in an atom-cavity system refers to a regime in which coherent interactions between the atoms and the cavity mode dominate over dissipative processes in some way.  There are several useful definitions of strong coupling, which are each relevant in different situations and for different tasks.  The simplest criteria is that the rate of coherent interactions --- $2g$ in the single-atom case and $\Omega$ in the multi-atom case --- should exceed the dissipative rates $\kappa$ and $\gamma$ (as well as other sources of inhomogeneous broadening).  These criteria define the resolved vacuum Rabi splitting regime.

\begin{figure}[!htb]
\centering
\includegraphics[width=3.3in, ]{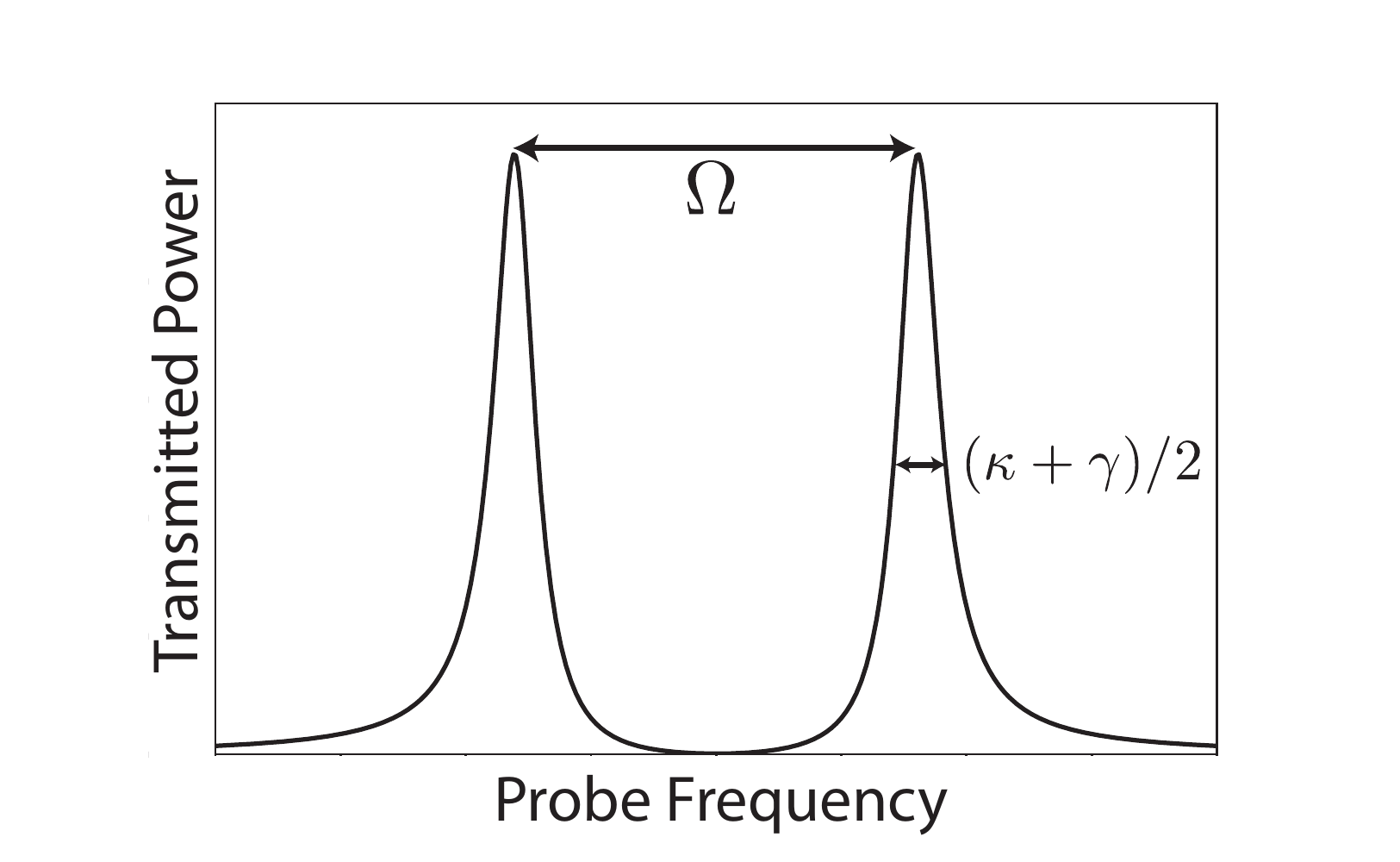}
\caption{A vacuum Rabi splitting for $\Omega \gg \kappa, \gamma$. A single transmission peak of the empty cavity splits into two features in the presence of atoms.  The splitting between these two peaks is set by $\Omega$, and the width of each peak is a weighted average of atomic and cavity decay rates.    }
\label{fig:vrs_thy}
\end{figure}

The eponymous signature of the resolved vacuum Rabi splitting regime ($\Omega \gg \kappa, \gamma$) can be observed by tuning the frequency of a weak probe tone incident on the cavity over a range surrounding the atomic and cavity transmission frequencies, and monitoring the power transmitted through the cavity (Fig.~\ref{fig:vrs_thy}).  When the cavity is tuned on resonance with the atomic transition ($\delta_c = 0$), a single transmission peak of the empty cavity splits into two peaks separated by $\Omega$, corresponding to hybrid atom-cavity modes centered at the eigen-frequencies of the collective Hamiltonian \cite{raizen1989normal}.  In the resonant case, the corresponding eigen-modes are half atom-like and half photon-like in character, and thus decay at a rate given by the average of the atom and cavity decays $\kappa^\prime = (\kappa + \gamma)/2$.  This decay rate determines the width of each of the two peaks.  
If the two peaks are resolved, it means that their separation, which is set by the rate of coherent interactions, exceeds their width, which is set by decay rates from the system.  

The single-atom resolved vacuum Rabi splitting regime is of particular interest in that it involves nonlinearities that occur between a single atom and a single photon.  This in turn enables applications such as single photon switches \cite{shomroni2014all, tiecke2014nanophotonic}, where one photon controls the state of another, and for deterministic single photon sources \cite{kuhn2002deterministic, mckeever2004deterministic}.

A second definition of the strong coupling regime
involves the cooperativity parameter, $\eta = 4 g^2/\kappa \gamma$.  In the limit that $\kappa \gg \gamma$, the cooperativity parameter quantifies the relative probability that an atom placed in the excited state in the cavity mode will decay by emitting a photon into the cavity mode, versus by emitting a photon into free space.  For many applications, $\eta \gg 1$ is a key figure of merit.  If many atoms are used, $\eta$ becomes collectively enhanced to a value of $N \eta$, which can greatly exceed unity.  This means that collective excitations are efficiently coupled to the cavity mode, from which they can be efficiently extracted and measured or transmitted.  

\noindent \subsection{Experimental regimes}
Strongly-coupled atom-cavity systems have typically relied on either microwave domain transitions between highly excited Rydberg states, or dipole allowed optical transitions.  

The first of these approaches involves coupling Rydberg atoms to a superconducting microwave cavity, as reviewed in ref.~\cite{raimond2001manipulating}.  Such cavities can have very low loss (small $\kappa$), enabling photon lifetimes of millisecond timescales.  Further, because transitions between Rydberg states can have large dipole moments (of order 1000 atomic units), they couple strongly to the microwave field (large $g$, typically several 10's of kHz).  Finally, because the only available decay channels are at microwave frequencies, the available phase-space for spontaneous emission is relatively small, which leads to a long radiative lifetime of up to several tens of milliseconds (small $\gamma$), though available interaction times in atomic beam experiments are typically much shorter.  These qualities, combined with efficient state-selective detection techniques, made this an ideal system for early explorations of the strong coupling regime, leading to a wealth of pioneering work \cite{raimond1982statistics, kaluzny1983observation, RevModPhys.73.565}.   


The second approach is to couple Alkali atoms to optical resonators using dipole-allowed optical transitions.  Certain physical scalings lead to significant challenges with this approach, especially in achieving single-atom strong coupling, but the allure of coupling atoms to optical fields (whose thermal occupation at room temperature is low \cite{kimble2008quantum}) has justified great efforts in this domain \cite{ritter2012elementary, boozer2007reversible}.  

In order to overcome the rapid ($2\pi \times$ several MHz typical) spontaneous emission associated with optical transitions (which scales as the cube of frequency), one must confine the cavity mode to very small volumes to increase the local electric field at the location of the atoms, and thus the rate of coherent interactions.  Even if one assumes that the mode volume scales with the wavelength of the radiation as $1/\omega^3$ (which practically becomes very hard to do), the ratio $g/\gamma$ scales unfavorably with frequency as $1/\omega$.  Counterintuitively, it is the relatively small dipole matrix element (typically around 5 atomic units for common species) of these optical transitions that makes them viable for strongly coupled applications at all --- $g$ scales with $\vec{d}$, while $\gamma$ scales with $\vec{d}^2$, so achieving $g/\gamma \gg 1$ is actually easier with a weak transition matrix element $\vec{d}$, though this does make it harder to achieve $g \gg \kappa$.  This provides a hint that transitions with even smaller dipole matrix elements may be favorable.  

Finally, in order to maximize $g$ (to achieve $g \gg \gamma$), systems that seek strong single-atom coupling must use very short optical cavities (sub-millimeter lengths) with very tightly focused waists in order to attain a small mode volume $V$.  Unfortunately, while $g \propto 1/\sqrt{L}$, $\kappa \propto 1/L$.  This means that the short cavity length $L$ makes it challenging to achieve the condition $g \gg \kappa$.  The solution to this is to engineer mirror coatings with exceedingly high reflectivity.  Breakthroughs in dielectric mirror coatings have enabled the realization of cavities with finesse $F$ exceeding $10^5$ \cite{rempe1992measurement}, and the attainment of strong coupling with single atoms with coupling rates of several tens of MHz compared to atomic and cavity decay rates of several MHz  \cite{miller2005trapped}.  

Of course, attaining strong coupling is much easier if one uses many atoms instead of a single atom.  In doing so, one can substantially enhance the rate of coherent interactions between the atoms and cavity.  However, one then loses the single-particle nonlinearities that motivate the single-atom strong coupling regime.  This makes collectively coupled systems more classical than those with single-atoms in that quantum fluctuations of a single particle have a smaller effect on the behavior of the system.  However, such systems can still be used to study quantum phenomena.  For example, collective interactions between atoms and a cavity can be used to generate entangled atomic states \cite{mcconnell2015entanglement, hosten2016measurement, cox2016deterministic, LSM10, barontini2015deterministic}, and to create entanglement between collective atomic excitations and single photons \cite{DLCZ01, simon2007interfacing, mcconnell2015entanglement}.  Even when operating in a regime where quantum effects are of little importance, systems exhibiting strong collective coupling can be both interesting and useful, for example enabling spatial self-organization \cite{BGB2010}, nondestructive measurement of atomic observables \cite{PhysRevA.89.043837}, and novel forms of laser \cite{MYC09}.

\noindent \subsection{Narrow linewidth transitions}

Recent work has enabled strong collective coupling of ensembles of atoms to optical cavities using narrow and ultra-narrow optical transitions present in alkaline earth (and alkaline earth-like) atoms \cite{PhysRevLett.114.093002, christensen2015nonlinear, braverman2019near, norcia2015strong, Laske2019, norcia2016superradiance}.  These transitions occur between spin singlet and spin triplet states of these two-valence-electron atoms.  While nominally forbidden by selection rules, spin-orbit and electron-nuclear spin coupling effects lead to a small but finite decay linewidth for certain transitions \cite{boyd2007nuclear}.  In particular, the $^1$S$_0$ to $^3$P$_1$ transition in calcium, strontium and ytterbium have linewidths of 375~Hz, 7.5~kHz, and 184~kHz, respectively.  This can be compared to typical transitions allowed by selection rules at similar wavelengths, which have linewidths of several MHz.  In addition to these narrow linewidth transtions, ultranarrow linewidth ``clock" transitions also exist between $^1$S$_0$ and $^3$P$_0$.  These transitions have linewidths on the mHz scale in fermionic isotopes, and even narrower for bosons.  

Despite their weak dipole matrix elements, it is possible to achieve strong collective coupling between these narrow linewidth transitions and an optical cavity.  For doing so, these week transitions have advantages but also present additional challenges.

While achieving the condition $\Omega \gg \gamma$ is easier for a narrow linewidth transition (because $\Omega$ scales less strongly with $d$ than does $\gamma$), sources of atomic broadening besides photon emission into free space can make the use of narrow linewidth transitions challenging.  Effects like Doppler shifts ($\Delta \omega_D = 2 \pi v/\lambda$, where $v$ is the atomic velocity and $\lambda$ is the wavelength of the radiation) do not scale with the strength of the atomic transition.  Ensuring that these effects do not overwhelm coherent interactions, which are now much slower, requires far more care and lower temperatures than for dipole-allowed transitions.  
Specifically, assuming a fixed wavelength, in order to fix the same ratio of $g/\Delta \omega_D$ between a narrow linewidth transition and a broad linewidth transition, one must scale the temperature by the ratio of the transition linewidths, or tightly confine the atoms to less than an optical wavelength.   

Further, achieving the condition that $\Omega \gg \kappa$ is also more challenging, and requires a narrow linewidth cavity.  This can be achieved by using a longer cavity, as the linewidth decreases more rapidly with the cavity length $L$ than does $g$.  This has the added advantage of enabling easier optical access to atoms within the cavity.  

For our second definition of strong collective coupling, $N\eta \gg 1$, the scalings work out somewhat differently.  Because $g \propto d$ and $\gamma \propto d^2$, the dipole matrix element actually cancels from $\eta$.  Thus, for this definition of strong coupling, the fact that we use dipole-forbidden transitions is fundamentally unimportant.  However, it is often important that $N\eta\gamma$, the rate of collectively enhanced emission per atom in an overdamped regime, exceeds the rate of inhomogeneous dephasing, which can be challenging for transitions with very narrow linewidths.

There is one more consideration in an atom-cavity system where the transition linewidth is critical: the ratio of $\gamma$ to $\kappa$.  The regime where $\kappa \gg \gamma$ is often referred to as the bad-cavity regime, while $\gamma \gg \kappa$ defines the good-cavity regime.  In the bad-cavity regime, excitations are more likely to leave the atom-cavity system as photons transmitted through the cavity mirrors, while in the good-cavity regime they are more likely to leave as spontaneously emitted photons.  In many applications, photons transmitted through the cavity mirrors are desirable, as they provide only collective information about the atomic ensemble and are easily collected and measured.  Photons scattered into free-space are typically detrimental, as they lead to dephasing of the atomic ensemble and are difficult to utilize.  This makes the bad-cavity regime desirable for many applications, and can therefore make the use of dipole-forbidden transitions advantageous.

\noindent\subsection{Classical model for atom-cavity system}

\begin{figure}[!htb]
\centering
\includegraphics[width=3.3in, ]{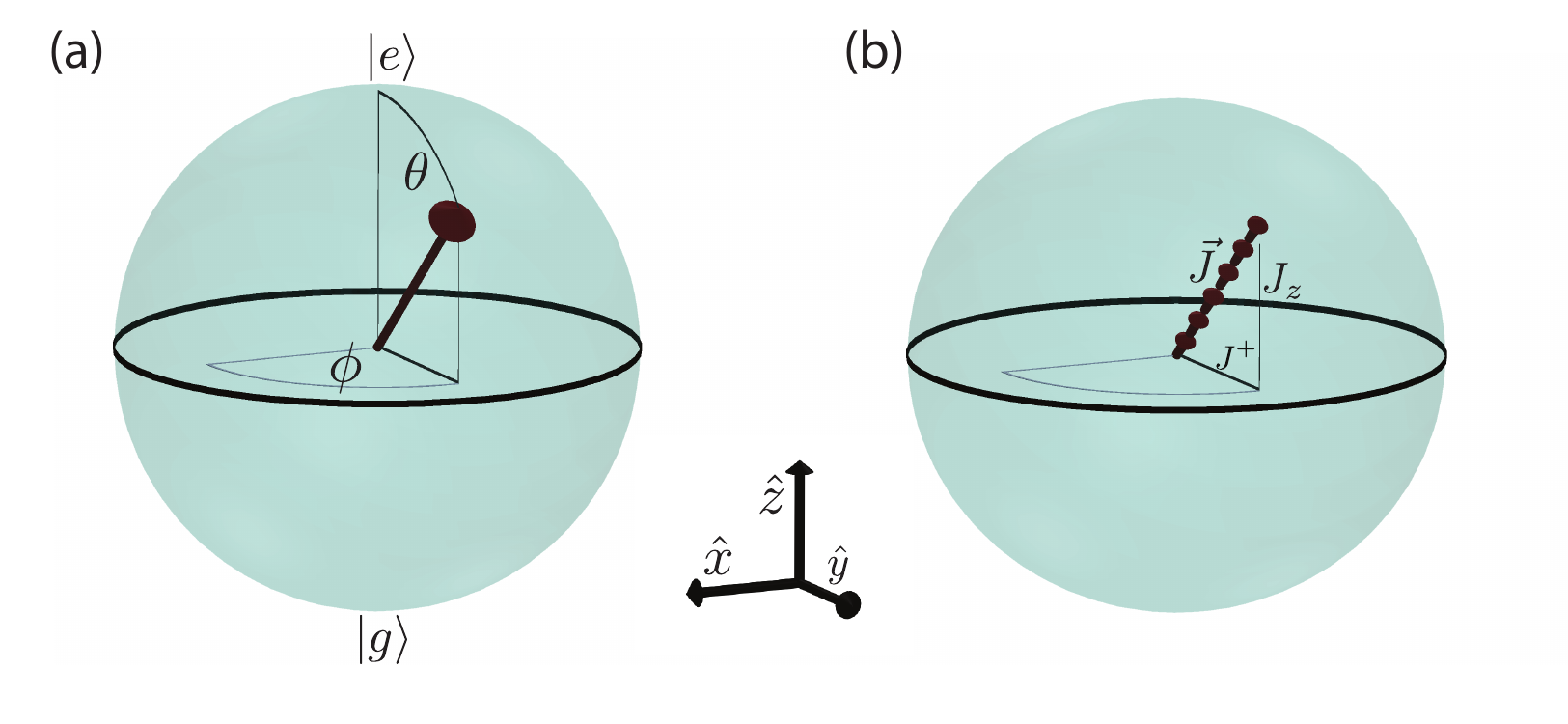}
\caption{
(a)   The state of a two-level system with states $\ket{e}$ and $\ket{g}$ can be represented by a Bloch vector on a Bloch sphere.  The figure shows a general state $\ket{\psi}=\mathrm{sin}(\theta/2)\ket{g} + \mathrm{cos}(\theta/2)e^{i \phi} \ket{e}$.  The vertical projection of the Bloch vector represents the population inversion of the state, while the azimuthal angle $\phi$ represents the relative phase between the $\ket{e}$ and $\ket{g}$ components.  (b) An ensemble of many two-level systems can be represented on a collective Bloch sphere by adding up the Bloch vectors associated with the individual two-level systems to form a collective Bloch vector $\vec{J}$.  The inversion of the collective Bloch vector is labelled $J_z$, and the atomic coherence (projection onto the equatorial plane) is labelled $J^+$.  }
\label{fig:BSBasics}
\end{figure}

The following section introduces a classical picture for describing the behavior of atoms interacting with an optical cavity, which is especially useful for gaining intuition into primarily classical, collective phenomena such as superradiance.  

We may describe a single atom as a two-level system, which can be represented by a vector on a Bloch sphere of radius $1/2$.  For pure quantum states, the tip of the vector lies on the surface of the sphere.  If the atom is in the excited state $\ket{e}$, the vector points to the north pole of the sphere.  If the atom is in the ground state $\ket{g}$, the vector points to the south pole.  An atom in the state $\ket{\psi}=\mathrm{sin}(\theta/2)\ket{g} + \mathrm{cos}(\theta/2)e^{i \phi} \ket{e}$ is depicted in Fig.~\ref{fig:BSBasics}a.

\begin{figure}[!htb]
\centering
\includegraphics[width=3.3in, ]{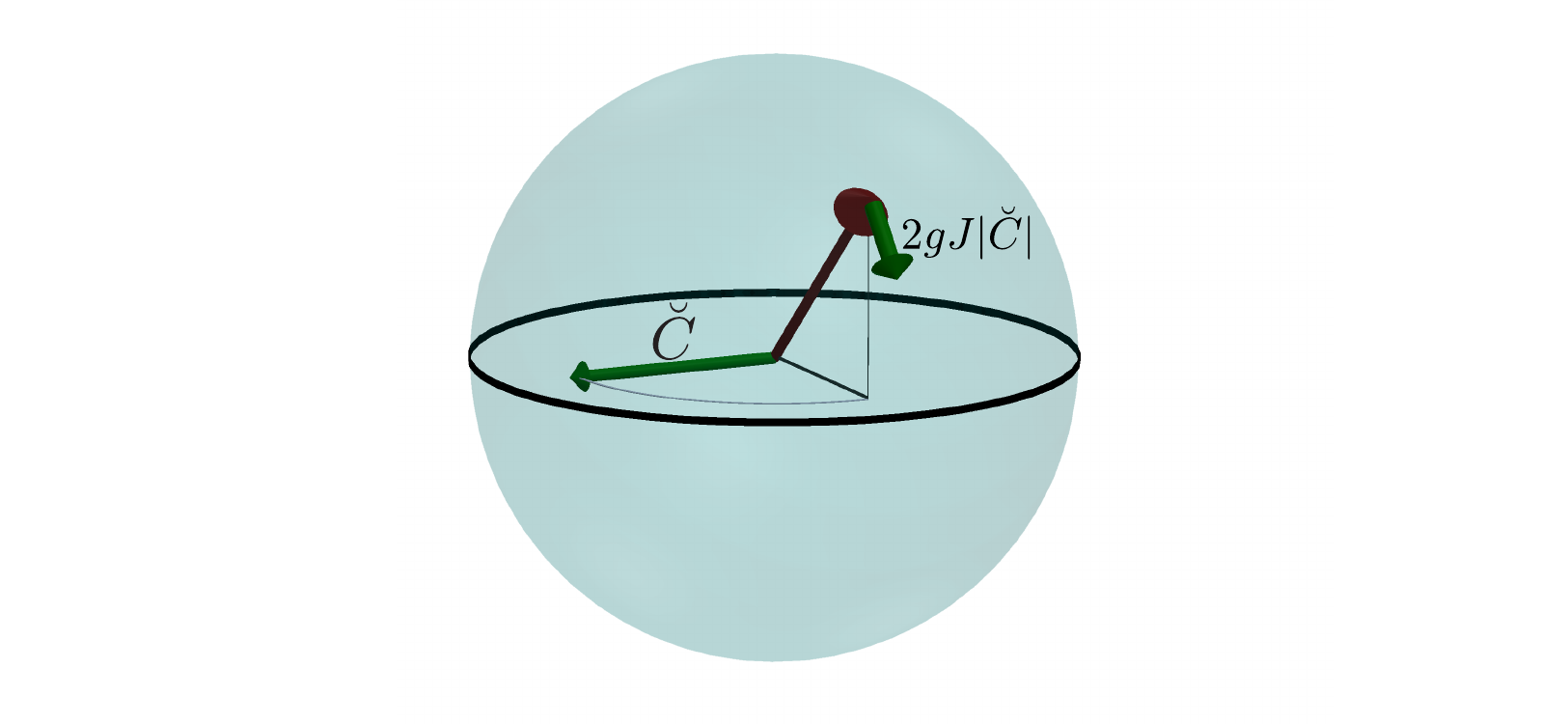}
\caption{Atomic dynamics in the presence of a cavity field $\breve{C}$, represented by an axis about which the Bloch vector rotates at a rate $2g|\breve{C}|$.  As pictured, the rotation drives the atoms from $\ket{e}$ to $\ket{g}$.  }
\label{fig:BSdyn}
\end{figure}

For systems with more than one atom, we simply add the Bloch vectors of the individual atoms together to form a collective Bloch vector $\vec J$, which resides on a collective Bloch sphere \footnote{In order to account for the fact that the atoms are located at different points along the cavity axis, and therefore experience a different optical phase from the cavity field, we define the phase of each atom's Bloch vector relative to the phase of the optical field of the resonant mode.  So long as the atoms do not move around, and coherent rotations are only performed using the cavity mode, we can safely ignore this detail.}.  We describe $\vec J$ in terms of it's vertical projection $J_z$, which is $-N/2$ at the south pole, 0 at the equator and $N/2$ at the north pole, and its projection onto the equatorial plane $J^+ = J_x + \imath J_y$, which is a complex number whose real and imaginary parts are $J_x = J \mathrm{sin}\theta\mathrm{cos}\phi$ and $J_y = J \mathrm{sin}\theta \mathrm{sin}\phi$ respectively.  These quantities are labelled in Fig.~\ref{fig:BSBasics}b.

We would like to describe the motion of the Bloch vector in the presence of a cavity field $C$, near resonance with an atomic transition.  
In general, $C$ can be any arbitrary cavity field, from any source.  Here though, I will specialize to the case where the cavity field is radiated by the atoms themselves, as this enables us to understand the origin of collective enhancement and other key aspects of superradiant emission.  
Following the notation of ref.~\cite{bohnet2014linear}, we may write down differential equations (known as optical Bloch equations) that describe the interactions with the cavity field $C$, as well as dissipation.  We assume that both the atomic coherence and cavity field oscillate at the same as of yet unknown frequency $\omega_\gamma$, and define $\breve{J}^+ = J^+ e^{\imath \omega_\gamma t}$ and $\breve{C} = C e^{\imath \omega_\gamma t}$.  We may then write equations of motion for $\breve{J}^+$ and $\breve{C}$:

\begin{eqnarray}
    \dot{\breve{J}}^+ &=& -[\gamma_\perp + \imath(\omega_a - \omega_\gamma)]\breve{J}^+ + \imath 2 g \breve{C}J_z \label{eq:Jdot}\\
    \dot{J_z} &=& -\gamma(J_z + N/2)/2 + i g (\breve{J}^+ \breve{C}^* - \breve{J}_-^*\breve{C})
    \label{eq:Jzdot}\\
    \dot{\breve{C}} &=& -[\kappa/2 + \imath(\omega_c - \omega_\gamma)]\breve{C} - \imath g \breve{J}^+
\end{eqnarray}

Here, $\gamma_\perp$ defines the decay rate of $J^+$ due to excited state decay or other processes.  In the absence of dephasing processes other than excited state decay, $\gamma_\perp = \gamma/2$.

We can represent these equations of motion on the Bloch sphere by associating with $\breve{C}$ a vector that originates from the center of the Bloch sphere. For a field whose frequency is on resonance with the atomic transition, this vector lies in the equatorial plane of the Bloch sphere.  The Bloch vector then rotates about this cavity field vector at a frequency $2g|\breve{C}|$, with $\breve{C}$ expressed in units of the amplitude associated with a single photon, as shown in Fig.~\ref{fig:BSdyn}. If $\breve{C}$ were constant (as would be the case for an externally applied drive, rather than a self-generated field) and in the absence of atomic damping, the Bloch vector would simply rotate at a constant rate about $\breve{C}$, as we would expect for simple Rabi precession.  In the case of a self-generated field however, $\breve{C}$ varies in time in a manner dependent on the atomic state, the cavity detuning $\delta_c$ and on the damping rate $\kappa$.  This dependence can lead to nontrivial atomic dynamics, and is critical for understanding phenomena such as superradiance.

\section{Applications}
The remainder if this tutorial describes several applications where coupling atoms to a cavity using narrow linewidth optical transitions can provide new opportunities for optical frequency metrology. Due to the author's area of expertise, these will focus in particular on work performed recently at JILA using strontium atoms.  While this is not intended to be a comprehensive review, brief descriptions of other relevant work will be provided when relevant.  

Broadly speaking, the goal of optical frequency metrology is to realize an oscillator operating in the optical domain -- a laser -- whose frequency is as stable as possible, and for many applications is also known in absolute terms.  The stabilized laser then defines a reference for time and distance that can then be utilized directly in the optical domain, or transferred to the microwave domain using a frequency comb \cite{diddams2000direct, fortier2011generation}.  Optical frequency metrology has led to significant advances in timekeeping by enabling the most precise and accurate clocks ever created \cite{schioppo2017ultrastable, Nicholson2015, origlia2018high, ushijima2015cryogenic}, which in turn allows applications to relativistic geodesy \cite{grotti2018geodesy}, tests of relativity \cite{sanner2018optical, chou2010}, and searches for dark matter \cite{arvanitaki2015searching}. 

One way to stabilize the frequency of a laser is to use electronic feedback to lock it to a resonance of a highly engineered optical cavity \cite{drever1983laser}.  Such techniques enable high-bandwidth feedback to the laser frequency, and thus excellent stability on short timescales \cite{matei20171}.  However, thermal and technical drifts in the length of the reference cavity lead to persistent challenges for stability on long timescales \cite{numata2004thermal, notcutt2006contribution,robinson2019crystalline}.  Further, as manufactured objects, each cavity has a slightly different resonance frequency, which is not inherently linked to a universally known quantity.  

To overcome these limitations, an optical cavity is often used in conjunction with an atomic reference -- defined by an optical atomic transition -- which provides a universal, unchanging frequency against which to stabilize the laser on long timescales.  There are several ways to implement such an atomic reference, each with different tradeoffs in terms of complexity, accuracy, precision, and bandwidth.  

A relatively simple approach, typically compatible with high bandwidth feedback, is to use perturbations in the amplitude or phase of a weak probe laser transmitted through an ensemble of atoms to stabilize the laser. Such techniques have formed the basis of critical laser frequency stabilization techniques used in laboratory applications and for many portable references.  
When the highest levels of absolute accuracy and precision are required, a favorable technique is to apply laser light to an ensemble of atoms near the frequency of an atomic transition and then measure the probability for the atoms to be transferred from one state to another. 
In addition to enabling long atom-light coherence times, such protocols encode the effects of the atom-light interaction on atomic populations, which can in turn be measured with high accuracy.  
These features make such passive atomic clocks well-suited for use in applications like primary frequency standards \cite{ludlow2015optical}.  Finally, one can create an active atomic clock \cite{CHE09, MYC09}, where the frequency of light emitted from an atomic transition is used to stabilize a laser.  This new technique holds promise for creating a frequency reference with advantages in bandwidth and dynamic range over passive devices, as frequency measurements can be performed rapidly and continuously without precise prior knowledge of the atomic transition frequency, which may be useful when operating outside of a carefully controlled laboratory environment.  The following sections of this tutorial describe ways in which each of these approaches can be enabled or improved by coupling atoms with to optical cavities using narrow linewidth optical transitions.

\subsection{Laser frequency stabilization using light transmitted through atomic ensembles}

Perhaps the simplest way to stabilize the frequency of a laser relative to an atomic transition is to observe perturbations to laser light transmitted through a gas of atoms when the frequency of the laser is near that of the atomic transition.  
In order to reduce one's sensitivity to technical drifts, it is advantageous to utilize a narrow spectroscopic feature, the width of which is ultimately limited by the decay rate of the atomic transition being used.  This makes narrow linewidth optical transitions like those in alkaline earth atoms appealing candidates for laser frequency stabilization.  Such narrow linewidth transitions have been successfully used for laser frequency stabilization by free-space interrogation of a thermal beam \cite{fox2012high}.
The downside of a narrow linewidth transition, however, is that it interacts relatively weakly with the probe light.  This limitation can be mitigated by enhancing the atom-cavity interaction using an optical cavity.  

One promising approach to cavity enhanced spectroscopy has been demonstrated using the 7.5~kHz linewidth $^1$S$_0$ to $^3$P$_1$ transition in an ensemble $^{88}$Sr atoms at roughly millikelvin temperature \cite{tieri2015laser, PhysRevLett.114.093002, christensen2015nonlinear}.  Interestingly, despite the several MHz broad Doppler width associated with this temperature, a spectroscopic feature of well below 100~kHz can be obtained with photon-shot-noise limited signal-to-noise compatible with laser stabilization to a linewidth of 500~mHz \cite{christensen2015nonlinear}.  With straightforward technical improvements, this limit could be improved to below 10~mHz \cite{tieri2015laser}, competitive with state-of-the-art optical cavities \cite{matei20171}.  Importantly, the experimental complexity of such systems is relatively moderate, potentially enabling deployment outside of research labs.  

\begin{figure}[!htb]
\includegraphics[width=3.375in, ]{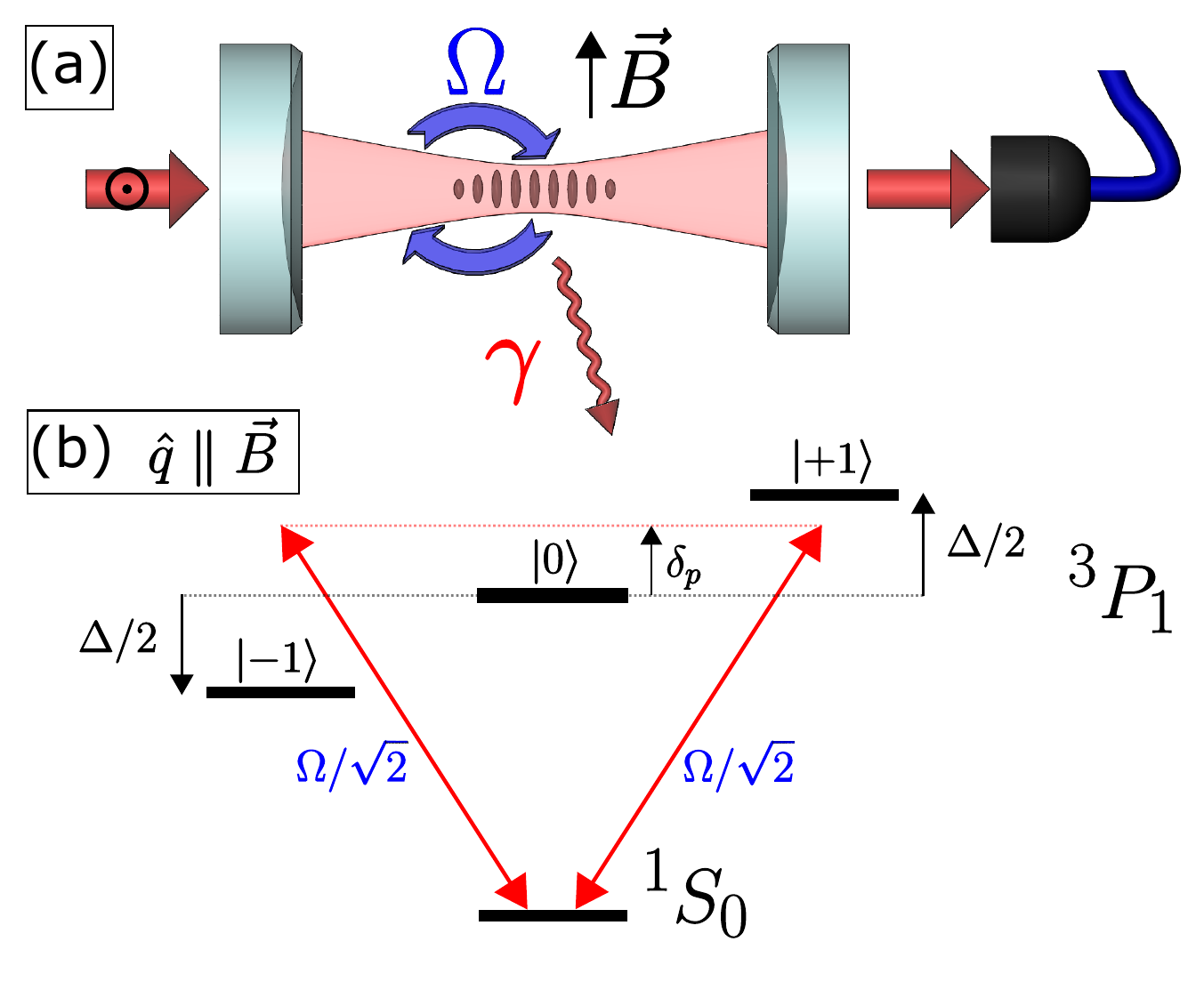}
\caption{(a)  The atom-cavity system is probed with light polarized perpendicular to an applied magnetic field. The light can be coherently absorbed by the atoms (brown ovals) and reemitted into the cavity at collective vacuum Rabi frequency $\Omega$. (b) The applied magnetic field creates a Zeeman splitting $\Delta$ between the excited states $\ket{\pm 1}$.  Transitions to each of these states interact equally with the probe light with collectively enhanced Rabi frequency $\Omega/\sqrt{2}$.  
The $\ket{0}$ state is shown, but it does not interact with the horizontally polarized cavity-field. Figure reproduced from ref.~\cite{winchester2017magnetically}.  
}
\label{fig:leveldiagram}
\end{figure}

\begin{figure}[!htb]
\includegraphics[width=3.375in, ]{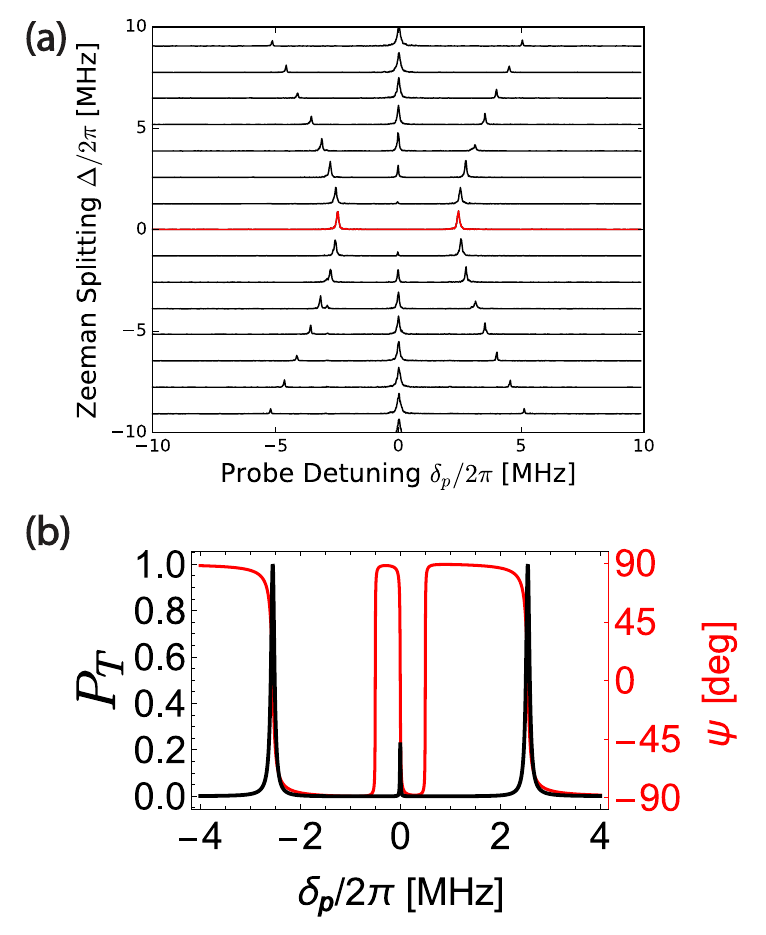}
\caption{(a) The transmitted power through the cavity versus the probe detuning $\delta_p$, with $\delta_c=0$. Each trace was taken for different applied magnetic fields, creating different Zeeman splittings $\Delta$ labeled on the vertical axis. The central red trace is taken for $\Delta=0$ and displays a collective vacuum Rabi splitting $\Omega/2\pi=5$~MHz.  When a magnetic field is applied perpendicular to the probe polarization, inducing a Zeeman splitting $\Delta$, a new transmission feature appears in between the two original resonances of the vacuum Rabi splitting.  
(b) Linearized theory showing the power $P_{T}$ and phase $\psi$ of the transmitted light, plotted here for $\Omega/2\pi=5$~MHz and $\Delta/2\pi=1$~MHz. Figure reproduced from ref.~\cite{winchester2017magnetically}.  }
\label{fig:waterfall}
\end{figure}

In an alternative configuration, referred to as magnetically induced transparency (MIT) in analogy with the much-studied phenomena of electromagnetically induced transparency (EIT)\cite{boller1991observation}, a similar spectroscopic feature can be generated by applying a magnetic field to an ensemble of $^{88}$Sr atoms coupled to a cavity, and probing the cavity with a laser polarized perpendicular to the magnetic field.  The following section provides a brief, qualitative picture of this effect.  For detailed and quantitative descriptions, see ref.~\cite{winchester2017magnetically}.  

In the MIT configuration, the probe light couples the ground state to two optically excited Zeeman sublevels, which are split by a frequency $\Delta$ in the presence of a magnetic field (fig.~\ref{fig:leveldiagram}).  When the cavity is placed at the frequency of the zero-field atomic transition, directly in between the two Zeeman sublevels, three transmission features can be observed -- one at the zero-field atomic transition frequency, one that is above the frequency of the upwards-shifted Zeeman sublevel and one that is below the frequency of the downwards-shifted Zeeman sublevel.  It is the first, central feature that is of interest for laser stabilization (fig.~\ref{fig:waterfall}a).  

This transmission feature results from a combination of two effects.  Firstly, the presence of the magnetic field shifts the two relevant atomic transitions away from the frequency of the cavity probe, creating a low-absorption frequency window through which the probe can pass.  Secondly, for transmission to be observed through the cavity, a resonance condition must be satisfied in which the round trip phase accrued by light circulating the cavity must be an integer multiple of $2\pi$.  Since the frequency of the spectroscopic transmission feature is nominally identical to that of the bare cavity, this implies that the atoms impart no net phase upon the cavity field.  This is indeed the case, as the probe light is above the frequency of one atomic transition and below the frequency of the other by the same amount, so the dispersive phase shifts cancel (fig.~\ref{fig:waterfall}b).  

Importantly, while the dispersive shifts imparted by the atoms cancel when the probe and cavity are both exactly on resonance with the zero-field atomic transition, the atoms still have important effects on the spectroscopic feature.  The presence of the atoms strongly influence the dispersion experienced by the probe light near the frequency of the spectroscopic feature.  For an atomic transition that is narrow relative to the optical cavity, this can lead to a substantial narrowing of the cavity transmission feature when atoms are present -- a useful feature for laser frequency stabilization.  Critically, in this regime where the atomic dispersion dominates over that of the cavity, the center frequency of the transmission feature is primarily determined by atomic transition frequency, rather than the resonance frequency of the cavity.  In this limit, the atoms provide a frequency-dependent shift to the phase of the probe laser on each round trip between the cavity mirrors.  The cavity then converts this phase shift into a change in output power, resulting in the narrow transmission feature.  

In ref.~\cite{winchester2017magnetically}, a spectroscopic feature with linewidth approaching the 7.5~kHz natural linewidth of the atomic transition was observed, confirming that the feature can indeed be made much narrower than the roughly 160~kHz linewidth of the optical cavity.  Further, a reduction in sensitivity of the spectroscopic feature to cavity frequency fluctuations of a factor of 20 was observed relative to an empty cavity.  This suppression, as well as the absolute linewidth of the spectroscopic feature, could be greatly improved by using a transition with narrower linewidth, such as the corresponding transition in calcium, with 375~Hz linewidth.  Importantly, while measurements in ref.~\cite{winchester2017magnetically} were performed with the atoms tightly confined in a lattice along the cavity axis, similar results were observed for unconfined atoms, which may enable application to a continuous beam in a relatively simple apparatus.

\subsection{Spin squeezing and nondestructive atom counting}

The coupling of ensembles of atoms to a cavity via a narrow linewidth optical transition holds promise for improving the performance of optical lattice clocks by facilitating the generation of entangled states with enhanced metrological performance, and through nondestructive readout of atomic populations.  In particular (as will be detailed in this section), the use of narrow linewidth transitions may provide certain practical advantages for the generation of metrologically useful entanglement on the optical transitions used in state-of-the-art atomic clocks.

Like many clocks, optical lattice clocks rely on state-dependent atom number measurements to infer frequency \cite{ludlow2015optical}.  Typically, these measurements are performed through fluorescence detection, where atom number is inferred by scattering light off of an atomic transition.  This technique has the advantage of simplicity and high signal-to-noise, but has the disadvantage that the scattered photons heat the atoms to the point where a new ensemble must be prepared.  The resulting dead-time required for atom preparation degrades the performance of the clock by leading to increased sensitivity to frequency noise on the laser used to interrogate the atoms \cite{Dick1987}.  

It is thus desirable to perform atomic population measurements while causing minimal heating of the atoms.  This can be accomplished in an optical clock either by observing phase shifts on a weak off-resonant probe that illuminates the atoms in a single pass \cite{LWL09}, or by observing perturbations to the resonances of a cavity mode containing the atoms.  In addition to enabling excellent precision, cavity-based measurements can be performed in a highly non-destructive manner.  

Collective interactions between atoms and a cavity can enable one to overcome another important limitation faced by optical lattice clocks, known as quantum projection noise \cite{Itano1993}.  
At the end of a clock sequence (but before the final measurement), atoms are typically left in a superposition of the two states that form the clock transition, with the clock signal encoded in the exact fraction that are in each of the two states.  When one measures this fraction, however, each atom projects into one state or the other, leading to fluctuations in the readout, or quantum projection noise.  This noise sets a limit on the clock resolution that scales inversely with the square root of the atom number.  However, if the atoms are prepared in a suitable entangled state, such as a ``squeezed" state, the random projections of individual atoms can be correlated, enabling measurements with greater precision than that set by the projection noise for independent atoms \cite{Wineland1992}.  
This squeezing can be generated in several ways, though I will focus here on those that rely on light-mediated interactions between atoms.  

One particularly effective protocol utilizes collective measurements of atomic state populations.  By performing suitably nondestructive population measurement both before and after the clock sequence and taking the difference between the two results, one can subtract out the quantum projection noise from the clock signal \cite{kuzmich98}. 
This procedure is referred to as measurement based spin squeezing, and relies on entanglement created between atoms by the collective measurement.  

Spin squeezing can also be performed using cavity feedback mechanisms that lead to entanglement between atoms without relying on measurement \cite{PhysRevA.81.021804, LSM10}.  In such protocols, the atomic state populations lead to tuning of a cavity mode, which in turn modifies the intensity of a probe laser applied off resonance from the cavity mode.  This probe laser causes light shifts to the atoms, and thus mediates collective interactions that can produce useful entanglement.    

Proof-of-principle demonstrations of spin squeezing have been performed using dipole allowed transitions in alkali atoms either in free space \cite{AWO09, kuzmich98, SKN12} or coupled to optical cavities \cite{hosten2016measurement, cox2016deterministic, LSM10, SLV10}.  An outstanding goal is to apply these techniques to a state-of-the-art optical lattice clock based on an ultra-narrow optical transition in an alkaline-earth or alkaline-earth-like atom.

The number of atoms in a particular atomic state, which I will now refer to as $N$, can be inferred by placing a resonant mode of a cavity that contains the atoms on or near resonance `with an atomic transition that involves that atomic state.  The collective interactions between the atoms and the cavity modify the resonance of the cavity mode in a manner that depends on $N$.  By applying a weak probe to the cavity resonance, one can infer by how much the resonance was perturbed , and therefore $N$. In order to create an entangled state useful for metrology, it is critical that the measurement provides only collective information about the state of the atomic ensemble -- it may reveal the total number of atoms $N$ in a given state, but not which atoms.  

One can either perform such measurements with the cavity mode on resonance with the atomic transition (the vacuum Rabi splitting limit), or in a far-detuned, dispersive regime.  In the resonant case, the presence of atoms (assuming strong collective coupling $\Omega = 2 g \sqrt{N} \gg \kappa, \gamma$) leads to a vacuum splitting whose size depends on the square root of the number of atoms $N$.  By measuring the size of this splitting, one can then infer $N$.  If the cavity is instead detuned from atomic resonance by an amount $\delta_c \gg \Omega$, the cavity mode shifts by an amount $g^2 N/\delta_c$.  This shift can also be measured in order to infer $N$.  

Fundamentally, one's ability to perform measurement-based spin squeezing is limited by the number of photons that must be scattered per atom, $m_s^\mathrm{proj}$, in order to resolve the atomic state populations at the projection noise level \cite{PhysRevA.89.043837}.  Because the scattering of probe photons reveals the state of individual atoms, it causes decoherence that degrades the resulting entangled state.  A small value of $m_s^\mathrm{proj}$ is thus favorable, as it indicates less decoherence for a given measurement precision.  

When probing with the cavity on resonance with the atomic transition, $m_s^\mathrm{proj} = \frac{1}{4 q N \eta}(1 + \gamma/\kappa)^2$, where $q$ quantifies the total efficiency of photon detection.  Because the cooperativity parameter $\eta$ is independent of transition linewidth, the transition linewidth is relevant only in the final term. This quantifies the relative likelihood that a photon leaves the system through a cavity mirror where it can be detected, versus being scattered into free space, which causes decoherence.  In this resonant regime, working with an atomic transition that is narrow relative to the cavity is advantageous.  

However, if one detunes the cavity far from resonance with the atomic transition, a favorable condition can be recovered even for a broad linewidth transition.  In a far-detuned regime, $m_s^\mathrm{proj} = \frac{1}{4 q N \eta}$, and the dependence on transition linewidth is no longer present.  

While there is no fundamental link between the strength of transition used to generate the spin squeezing and the character of the transition that separates the clock states (which sets the fundamental performance of the clock), the use of narrow linewidth optical transitions may have practical advantages for generating squeezing in optical clocks.  

When the linewidth of the probed transition is much narrower than that of the cavity, one can work on resonance with fundamental performance that approaches that achieved when working far off resonance with a broad transition.   Thus, the choice of squeezing transition must be made based on technical considerations such as the availability of low-loss mirrors and low-noise lasers at the wavelengths of the different atomic transitions.  

\begin{figure}[!htb]
\includegraphics[width=3.375in]{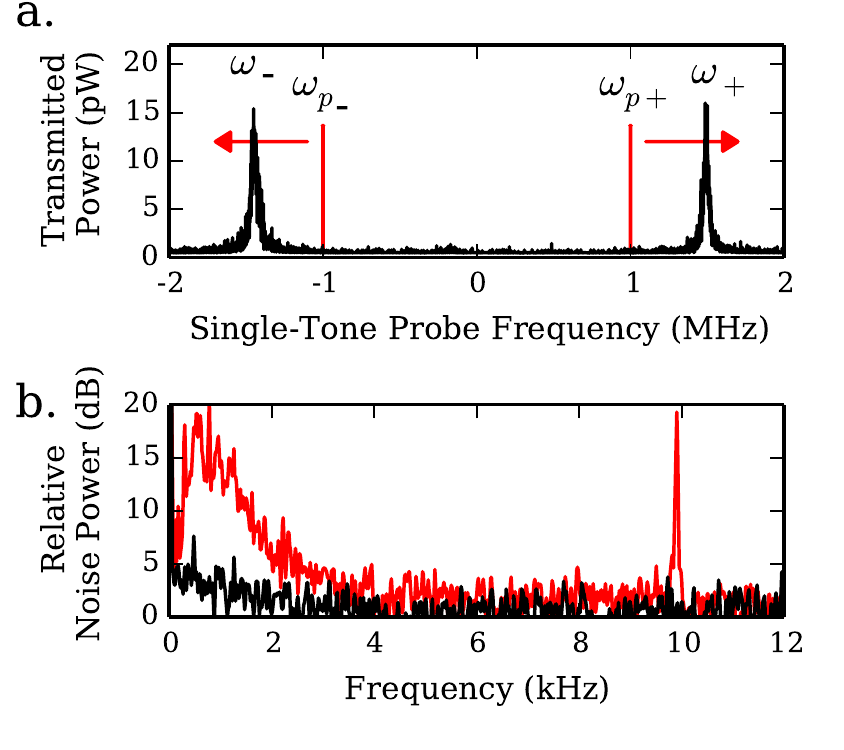}
\caption{(a) To probe the Vacuum Rabi Splitting, two probe tones at $\omega_{p\pm}$ are simultaneously swept across the two normal mode resonances at $\omega_{\pm}$ while total transmitted power is recorded.  (b) The noise power spectra relative to photon shot noise for cavity probe with a single tone (red trace) and two tones (black trace), demonstrate noise cancellation of two-tone probing. (figure reproduced from \cite{norcia2015strong})}
\label{fig:twotone}
\end{figure}

Working on resonance in the vacuum Rabi splitting limit also enables one to gain insensitivity to technical sources of noise by simultaneously probing both transmission features of the vacuum Rabi splitting \cite{CBS11} (see fig.~\ref{fig:twotone}a). 
In ref \cite{norcia2015strong}, such a probing technique enabled the demonstration of cavity enhanced atom counting with levels of precision and scattering compatible with the generation of spin squeezed states.  
In this work, a mode of a high-finesse optical cavity (linewidth $\kappa \simeq 160$~kHz) was placed on resonance with the 7.5~kHz linewidth $^1$S$_0$ to $^3$P$_1$ transition in $^{88}$Sr.  For an atom number of $N = 1.25 \times 10^5$, the observed vacuum Rabi splitting was roughly 5~MHz, placing the system deep in the desirable strong collective coupling, bad cavity regime $\Omega \gg \kappa \gg \gamma$.  

Small shifts in $\Omega$ were resolved by sweeping two probe tones simultaneously outwards in frequency across the transmission features of the vacuum Rabi splitting while monitoring transmitted power, as shown in fig.~\ref{fig:twotone}a.  When the probe tones are on the side of the resonance feature, the transmitted power is sensitive to changes in the atom number, which modify the size of the vacuum Rabi splitting, but insensitive to changes in the frequency of the cavity or probe laser, which move the two peaks in the same direction.  This insensitivity to laser frequency noise is shown in fig.~\ref{fig:twotone}b, and allowed measurement noise compatible with sub-projection noise readout even with a rather crude laser system (in this work, the atoms remained in the ground state, but the measurement precision was compared to the level of fluctuation that would be present if they were in an equal superposition state).  

Recently, cavity enhanced measurement combined with cavity feedback using the 184~kHz linewidth transition in $^{171}$Yb has enabled the observation of spin squeezing between the $^1$S$_0$ ground state Zeeman sublevels \cite{braverman2019near}.

In principle, this squeezing could be transferred to the metrologically relevant $^1$S$_0$ to $^3$P$_0$ optical clock transition, which would enable enhanced performance of an optical clock.  

In order to enhance the performance of an optical clock, entanglement must be present on the metrologically relevant $^1$S$_0$ to $^3$P$_0$ optical clock transition.  In the case of ref.~\cite{braverman2019near}, the entanglement demonstrated between nuclear spin states could in principle be transferred directly to the optical transition through the application of an optical rotation from one of the two entangled ground states to the optically excited $^3$P$_0$ state.  In order to generate spin squeezed states using the nondestructive measurements presented in ref.~\cite{norcia2015strong}, the measurements must be applied in conjunction with high-fidelity rotations on the optical clock transition.  Thus, both of these methods require the ability to perform precise optical rotations on the clock transition, which requires tight spatial confinement of the atoms along the direction from which the clock laser is applied in order to eliminate the effects of Doppler shifts and photon recoil.  This requirement presents challenges in experiments like those of refs.~\cite{braverman2019near, norcia2015strong}, where the cavity mirrors obscure optical access along the direction of tight atomic confinement, though this limitation could be overcome by additionally confining the atoms along a second direction.  

Cavity-mediated interactions and measurements using narrow linewidth optical transitions thus appear to be a promising path to creating useful entanglement on the optical transitions used in state-of-the-art clocks, and for rapid nondestructive measurements that would reduce clock dead-time.  These achievements could in turn mitigate two of the leading limitations on clock stability ---  atom projection noise and laser noise aliasing --- and could thus substantially improve the performance of state-of-the-art frequency metrology.  

\subsection{Superradiant frequency references}
Coupling atoms to a cavity using narrow linewidth transitions also opens up possibilities for a new form of active optical frequency reference based on superradiance \cite{MEH10, MYC09, CHE09}.  Superradiance refers to a phenomena in which the rate that photons are emitted from an ensemble of atoms is collectively enhanced, and greatly exceeds the sum of the rates at which photons would be emitted from the individual atoms if they were to radiate independently.  Superradiance itself has been studied in detail as a curiosity for decades \cite{DIC54, gross1982superradiance, skribanowitz1973observation, kaluzny1983observation}.  More recently it has been proposed \cite{MEH10, MYC09} that one could take advantage of an ultra-narrow linewidth clock transition like the one in strontium to realize a superradiant laser with linewidth of order 1 mHz.

Like any laser, a superradiant laser consists of a gain medium that sits within the mode of an optical buildup cavity.  In contrast to a conventional laser, where the gain medium provides broad-band amplification of power and the laser frequency is set by the resonance frequency of the optical cavity, a superradiant laser operates in a regime where the gain medium is by far the most spectrally selective element in the system, and the optical cavity serves to enhance emission in a comparatively broadband manner.  
The output frequency of the laser is then determined by the gain medium.  Because cold atoms with narrow linewidth optical transitions form a very narrow linewidth and stable gain medium, this could be a major advantage, potentially enabling one to overcome limitations associated with thermal and technical fluctuations associated with optical cavities \cite{numata2004thermal, notcutt2006contribution, robinson2019crystalline}.   
Further, because the light emitted from a superradiant laser comes directly from a narrow-linewidth atomic transition, one can view these devices as active clocks whose ticking rate is linked to atomic properties.  

Proof-of-principle demonstrations of a superradiant laser have been performed using Raman transitions in rubidium atoms, where a ``dressing" laser is used to create an effective long-lived optically excited state \cite{bohnet2012steady, bohnet2014linear, BCWHybrid, BCWDyn}.  While not useful as an optical frequency reference (the frequency noise on the dressing laser also appears on the laser output), this system was an ideal test-bed for studying many aspects of steady-state superradiance.  More recently, experimental explorations have been performed using true narrow linewidth optical transitions in strontium \cite{schaffer2019lasing, norcia2015cold, norcia2016superradiance, norcia2018cavity, norcia2018frequency} and calcium \cite{Laske2019}, which are compatible with the creation of a high-precision optical frequency reference.

\subsubsection{Collective enhancement of emission}
The mechanism behind collectively enhanced emission (superradiance) can be illustrated using the Bloch sphere and optical Bloch equations.  For the following, I will refer to the amplitude of the atomic coherence as $J_\perp$, with $J_\perp^2 = |\vec{J}^+|^2$. The superradiant regime is defined by an over-damped condition where $\kappa \gg \Omega, \gamma$, implying that the cavity field damps quickly compared to atomic dynamics.  For a cavity on resonance with the atomic transition ($\omega_a = \omega_c = \omega_\gamma$) we can thus consider quasi-steady state dynamics determined by:
\begin{eqnarray}
\dot{J}_z = - \eta \gamma J_{\perp}^2 - \gamma_\ell J_z,\\
\dot{J}_{\perp} = (\eta\gamma J_z -\gamma_{\perp} -\gamma_\ell)J_{\perp},\\
\breve{C} = - 2 i g J_\perp/\kappa.\label{eqn:mHzmeanfield}
\end{eqnarray}
Here, $\gamma_\ell$ represents atom loss, which for the very narrow linewidth transitions considered here is typically much faster than single particle spontaneous emission, and so terms representing single-particle decay have been omitted.  From these equations, we can understand the origin of collective enhancement in two ways: the rate of population decay ($\dot{J}_z$) has a term proportional to the square of the atomic coherence, and thus to the square of the number of atoms.  Also, the cavity field $\breve{C}$ is proportional to the atomic coherence, and because the power radiated from the cavity is proportional to $\breve{C}^2$, the output power is ultimately determined by the square of the number of atoms.  For maximally coherent atoms (described by a Bloch vector of length $J=N/2$ pointing along the equator) this leads to a photon output rate from the cavity of  $\frac{1}{4}N^2 \eta \gamma$.  

\begin{figure}[!h]
\centering
\includegraphics[width = 3.4in]{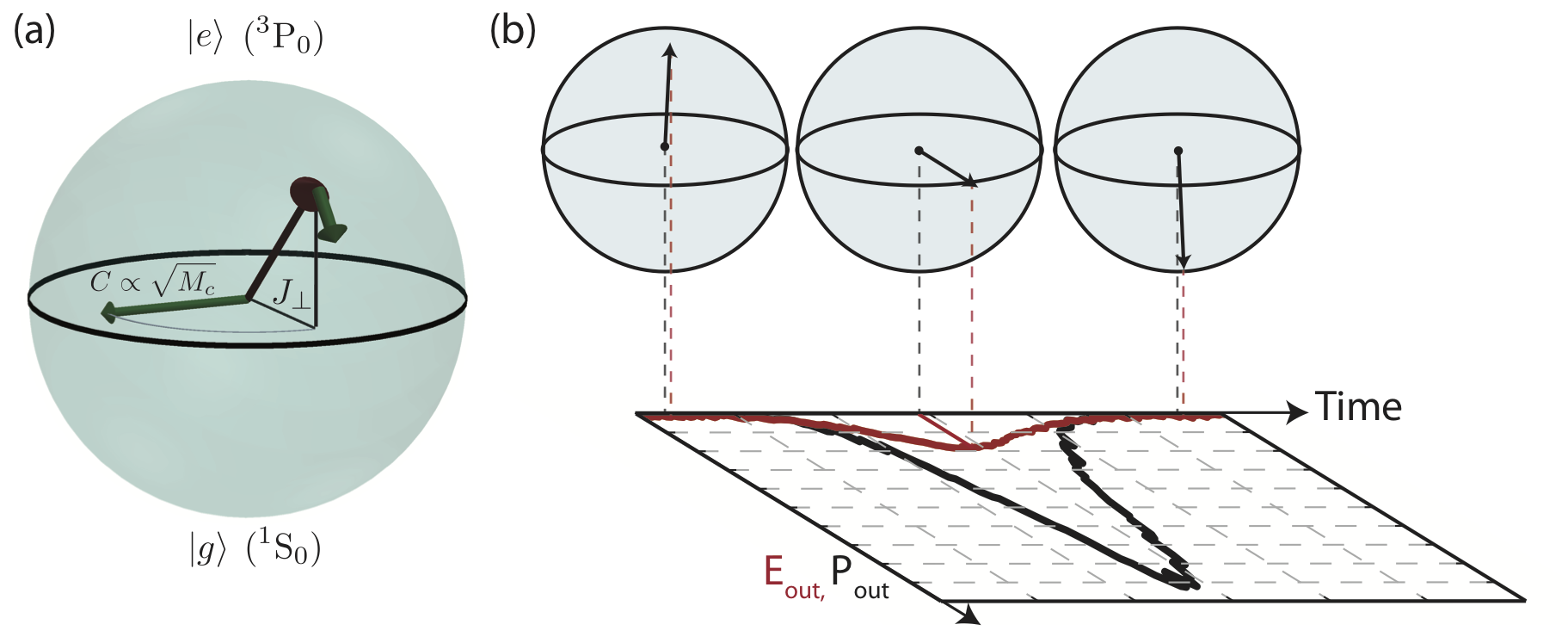}
\caption{(a) The atoms can be represented by a collective Bloch vector on a Bloch sphere.  The cavity field $\breve{C}\propto \sqrt{M_c}$, where $M_c$ is the number of photons in the cavity, is represented by a vector that lies in the equatorial plane perpendicular to the atomic coherence $J_\perp$ (when the cavity is on resonance with the atomic transition).  The Bloch vector rotates about the axis that represents the cavity field.  
(b) The Bloch vector behaves like a highly damped pendulum that starts inverted at the north pole of the Bloch sphere (all atoms in $\ket{e}$).  Quantum fluctuations disturb the system from its unstable equilibrium position, causing the Bloch vector to swing down the Bloch sphere, emitting peak radiation at the equator and ultimately relaxing to the south pole (all atoms in $\ket{g}$) as inversion is lost.  The radiated electric field (red trace) is proportional to the perpendicular projection of the Bloch vector, $J_\perp$, which at its peak is proportional to $N$.  The radiated power (black trace) is proportional to the square of the radiated electric field, and at its peak is therefore proportional to $N^2$.  Figure reproduced from ref.~\cite{norcia2016superradiance}.  
}
\label{fig:ExpDiag}
\end{figure}

For an ensemble whose Bloch vector lies in the northern hemisphere of the Bloch sphere, the atomic ensemble experiences positive feedback for emission.  The atoms radiate into the cavity, and the radiated electric field causes the Bloch vector to tip further from the north pole, increasing $J_\perp$ and thus causing the atoms to radiate more strongly.  The result is a pulse of light that builds up gradually, reaches a peak in power as the Bloch vector passes the equator, and falls to zero as the atoms reach the ground state (Fig.~\ref{fig:ExpDiag}b).

\begin{figure}[!htb]
\centering
\includegraphics[width=3.4in]{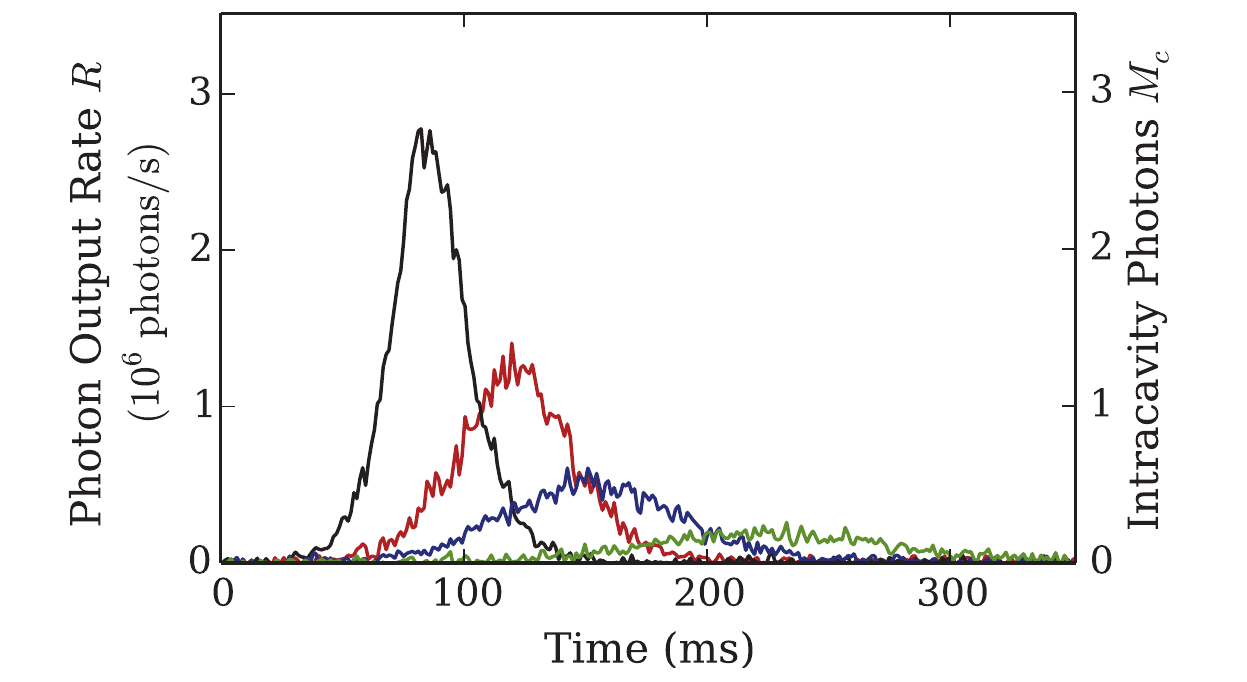}
\caption{Spontaneously generated superradiant pulses.  Representative single time traces of photon output rate $R$ for pulses at different atom number $N\approx$ $ 100\times10^3$ (green), $ 125 \times 10^3$ (blue), $150 \times 10^3$ (red), $ 200 \times 10^3$ (black).  The equivalent average intracavity photon number is calculated on the right as $M_c= R/\kappa$.  Figure reproduced from ref.~\cite{norcia2016superradiance}.  
}
\label{fig:mHzpulses}
\end{figure}

In ref.~\cite{norcia2016superradiance}, collectively enhanced pulses were observed from the mHz linewidth $^3$P$_0$ to $^1$S$_0$ clock transition in $^{87}$Sr by preparing atoms in the excited $^3$P$_0$, $m_f = 9/2$ state and measuring power emitted from the cavity (fig.~\ref{fig:mHzpulses}).  Pulses recorded with different numbers of atoms exhibit clear signatures of collective enhancement:  for larger atom numbers, the pulses are shorter in duration, appear sooner, and have a higher peak power, scaling approximately quadratically with atom number.  In practice, analysis is complicated slightly by the presence of atom loss, decoherence and inhomogeneous coupling of the atoms to the standing-wave cavity mode.  These effects are treated in detail in refs.~\cite{norcia2016superradiance, norcia2017new} and associated supplemental material.  

In general, it is important to consider the effects of tuning the cavity off resonance with the atomic transition, as this can perturb the frequency of the light emitted from the cavity.  In fact, when the cavity is detuned from atomic resonance, new dynamics emerge that lead to a frequency shift of the atomic transition that depends both on the atomic inversion and on the cavity frequency.  This causes the frequency of light emitted from the cavity to be shifted -- an effect known as ``cavity pulling." Such shifts also lead to atom-atom correlations via an effect known as ``one-axis-twisting" (OAT).  OAT has been demonstrated in driven systems as a way to create entangled spin-squeezed states \cite{LSM10,Leroux10, PhysRevA.81.021804,  bohnet2016quantum}.  In a superradiantly emitting system, the atoms themselves create the drive that leads to twisting, which may enable a new path to useful entangled states \cite{hu2017vacuum, lewis2018robust}.  These dynamics also lead to a spin-locking effect \cite{norcia2018cavity, davis2019photon}, similar to those observed in atomic systems with collision-mediated interactions \cite{allred2002high, McGuirk2002,Deutsch2010, Du2008}.

\begin{figure}[!h]
\centering
\includegraphics[width = 3.4in]{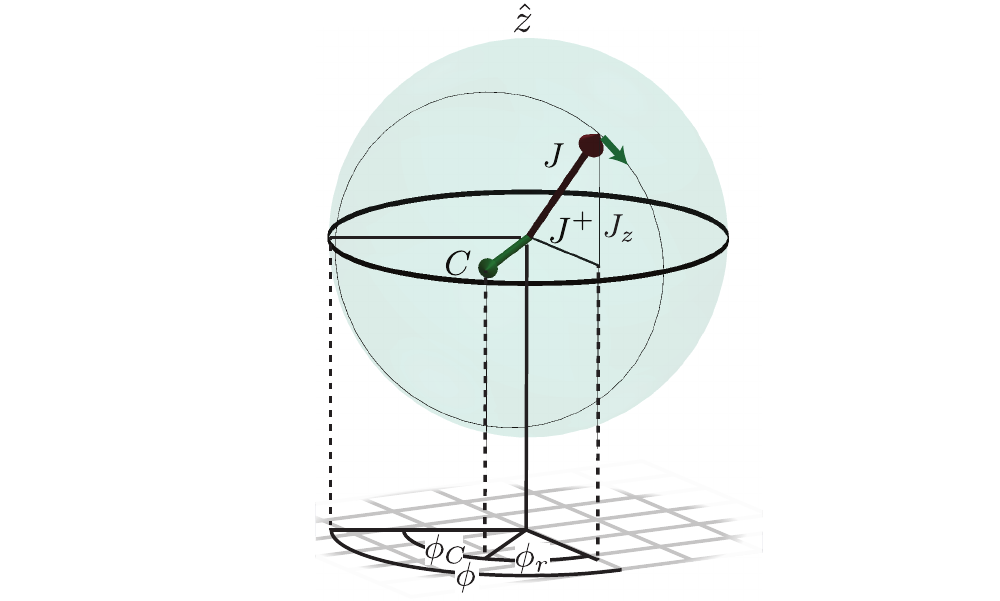}
\caption{A Bloch sphere representation of the atomic ensemble radiating into a detuned cavity.  The phase of the cavity field $\breve{C}$ ($\phi_C$) is locked to the phase of the atomic coherence $J^+$ ($\phi$) up to a fixed offset $\phi_r=\arctan(2\delta_c/\kappa)+\pi/2$.  The Bloch vector precesses about the cavity field axis, which is the source of both collectively enhanced decay and of shifts in the output frequency $\omega_\ell$.  
}
\label{fig:labels}
\end{figure}

\begin{figure}[!h]
\centering

\includegraphics[width = 3.4in]{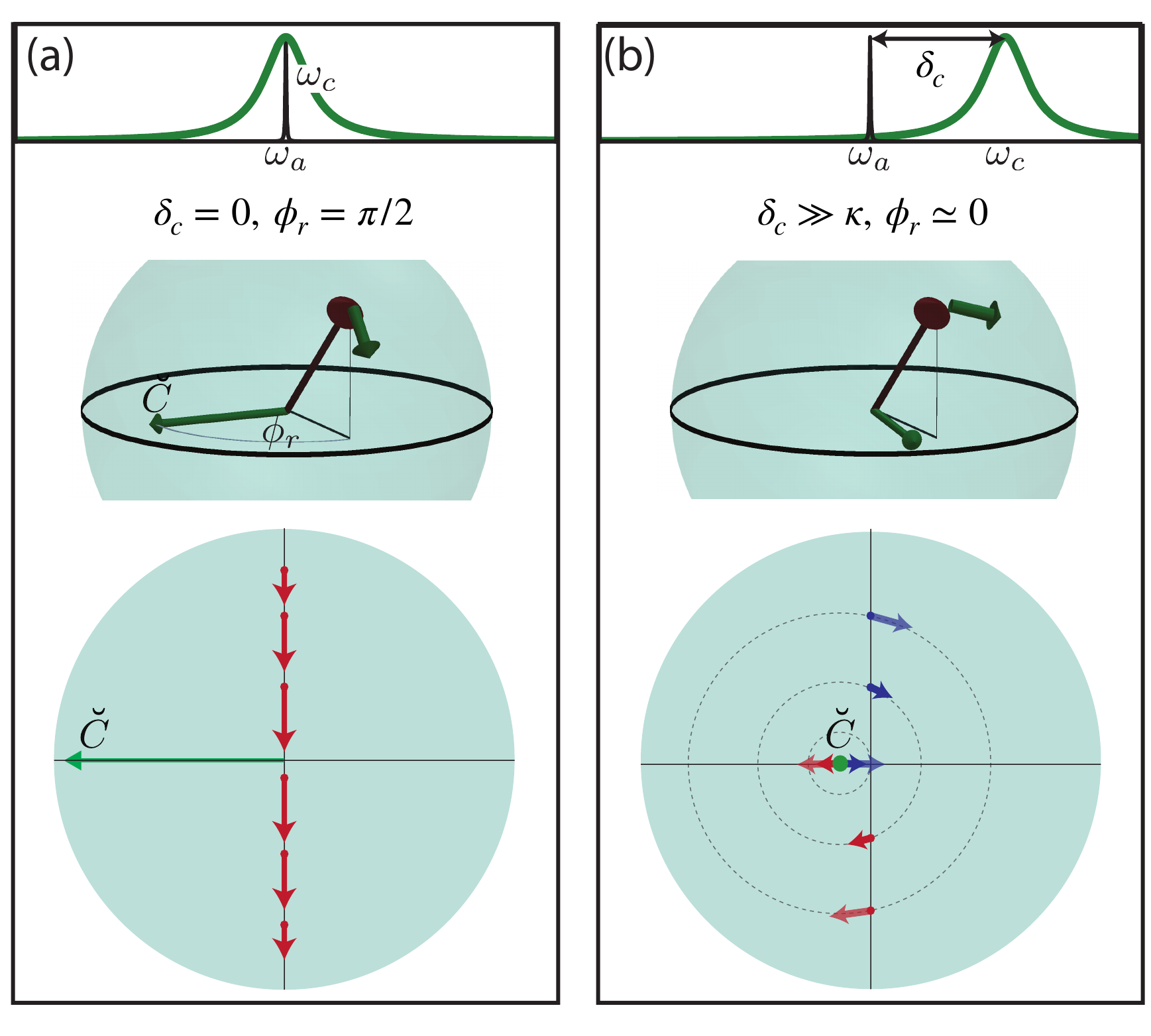}
\caption{(a) When the cavity is on resonance with the atomic transition ($\delta_c = 0$) the cavity field $\breve{C}$ is perpendicular to $J^+$.  This causes a rotation of the Bloch vector downwards, and is the physical origin of superradiant emission.  (b) When the cavity is tuned far off resonance from the atomic transition ($\delta_c \gg \kappa$), the cavity field $\breve{C}$ is nearly aligned with $J^+$.  Now, the rotation of the Bloch vector about $\breve{C}$ is primarily sideways.  Because $\breve{C}$ follows the Bloch vector as it rotates, this leads to a frequency shift of the Bloch vector precession, and of the emitted light.    
}
\label{fig:twobss}
\end{figure}

We can understnd the origin of cavity pulling using the Bloch sphere picture.  When the cavity is on resonance with the atomic transition ($\delta_c = 0$), the cavity field vector is always perpendicular to the atomic coherence.  This causes the Bloch vector to rotate downwards, and is the origin of the collectively enhanced emission (Fig.~\ref{fig:twobss}a).  In general, the collective emission leads to a decay of the collective excitation at a rate $\Gamma = 4g^2\kappa/(4\delta_c^2 + \kappa^2)$.  
When $\delta_c$ is non-zero, the phase between the cavity field $\breve{C}$ and the atomic coherence $J_+$ is modified.  This is because the atoms are now driving the cavity (a harmonic oscillator) off resonance.  Like any harmonic oscillator, the response of the cavity to an off-resonant drive will be shifted in phase relative to its resonant response --- it will exhibit a phase lag if the cavity is driven above resonance and a phase advance if it is driven below resonance.  On the Bloch sphere, this manifests as a phase shift of the rotation axis $\breve{C}$.   The field's azimuthal angle $\phi_C$ follows the azimuthal angle of the Bloch vector's projection onto the x-y plane $\phi$.  The relative azimuthal angle is $\phi_r=\phi-\phi_C$, and is set by the cavity detuning $\phi_r=\arctan(2\delta_c/\kappa)+\pi/2$.  

When the cavity is tuned far from resonance the field becomes nearly parallel or antiparallel with the atomic coherence ($\phi_r =0$ or $\pi$), depending on the sign of the detuning.  The effect of the field-generated rotation on the Bloch vector is no longer a rotation downward, but sideways --- the cavity field drives a rotation of the Bloch vector that changes its azimuthal angle, but not its polar angle (See Fig.~\ref{fig:twobss}b).  We interpret this additional precession of the Bloch vector's azimuthal angle as an inversion-dependent frequency shift $\omega_{OAT}=-2 \chi J_z$ of the atomic coherence $ {J}^+\sim{J}^+(0) e^{\imath\omega_{OAT} t}$, where $\chi = 4g^2\delta_c/(4\delta_c^2 + \kappa^2)$.  Because the cavity mode decays quickly, the cavity field vector will equilibrate to the new phase of the Bloch vector, maintaining the same relative phase $\phi_r$.  The net effect of this phase evolution is a frequency shift of both the atomic transition and the light emitted from the cavity.

In general, the dynamics will have contributions from both superradiant decay and cavity pulling, with classical equations of motion for the Bloch vector given by: 
 \begin{eqnarray}
 \frac{d J^+}{dt} &=&{-\rm i}({2\chi +{\rm i}\Gamma}) J^+(t) J_z(t) \label{eqn:MF}\\
  \frac{d J_z}{dt} &=&  { - \Gamma} J^+(t) J^-(t)\label{eqn:MFs}
\end{eqnarray}

In the absence of collective decay ($\Gamma = 0$), and assuming homogeneous coupling between the atoms and the cavity, the atomic system evolves according to: 
\begin{equation}
 \frac{dJ^+}{dt} = -2i\chi J_z(0) J^+ .\label{eqn:MF}
\end{equation}
\noindent Because $J_z$ is constant, this simply represents an azimuthal precession of the Bloch vector at a rate $2\chi J_z(0)$.

In order to experimentally observe the inversion dependence of cavity pulling, we prepare the atoms by performing a coherent rotation from the bottom of the Bloch sphere through an angle $\theta$ using resonant light coupled through the cavity.  Experimentally, the inhomogeneous coupling of the atoms to the cavity leads to some complications, described in detail in ref.~\cite{norcia2018cavity} and associated supplement.  In short, these can be treated by defining an effective coupling $g' = g/\sqrt{2}$ and an effective inversion $J_z' = N (\frac{J_1(\theta)}{\theta}-J_2(\theta))$, which reproduce the behavior of homogeneous coupling.

\begin{figure}[!h]
\centering
\includegraphics[width = 3.4in]{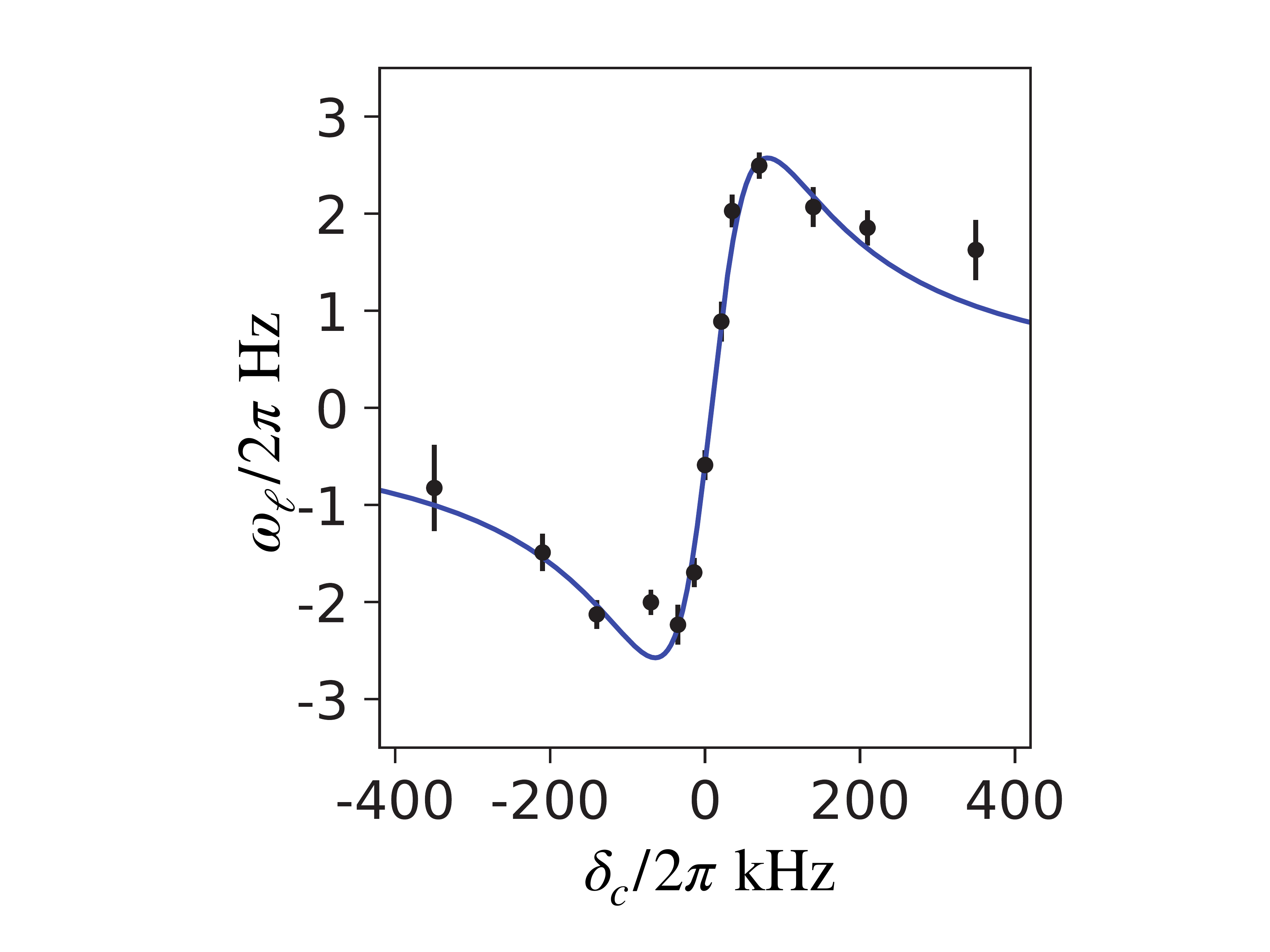}
\caption{Shifts in frequency of emitted light $\omega_\ell$ versus cavity detuning $\delta_c$, showing expected dispersive behavior.  Blue line is a fit with cavity linewidth held to its independently measured value. Figure reproduced from ref.~\cite{norcia2018cavity}.   
}
\label{fig:oat_exp_disp}
\end{figure}

Experimentally, we explore the variation of the frequency of emitted light $\omega_\ell$ with cavity detuning $\delta_c$ (Fig.~\ref{fig:oat_exp_disp}) at a given inversion. On each trial, we prepare the atoms in the same state near the bottom of the Bloch sphere (where the shifts are large), and scan the detuning of the cavity mode while recording changes in $\omega_\ell$.  $\omega_\ell$ is measured relative to a stable reference laser using a heterodyne beat note.  We observe the expected dispersive behavior of the frequency shift versus detuning, as can be seen by the fitted dispersive curve  with the cavity linewidth held fixed to its independently measured value. In the context of superradiant frequency references, it is the central slope of this feature that concerns us, as it represents the sensitivity of the output light to small perturbations in cavity frequency.   We can define the pulling coefficient $P = \frac{d \omega_\ell}{d \delta_c}$ to represent this sensitivity.  

\begin{figure}[!h]
\centering
\includegraphics[width = 3.4in]{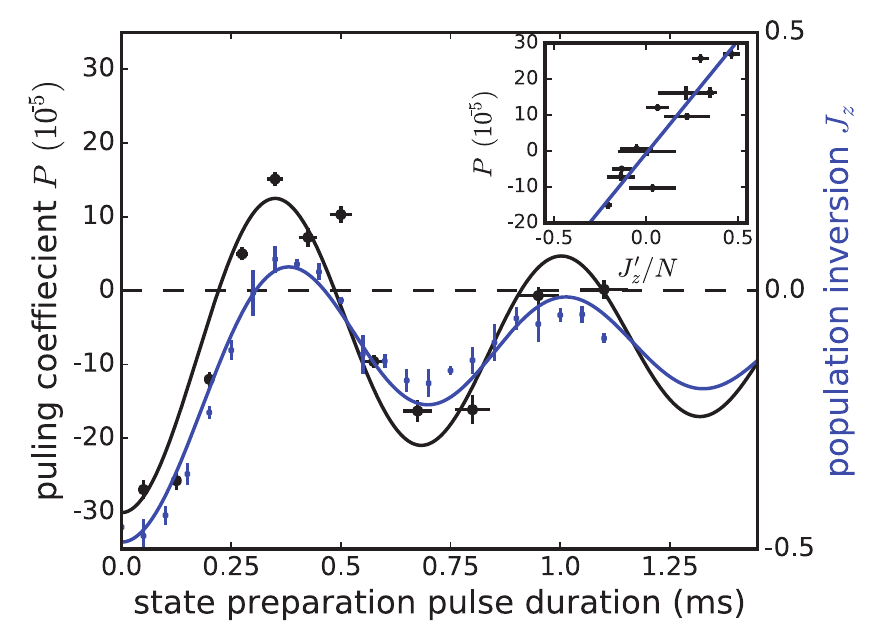}
\caption{Main: Pulling coefficient $P$ (black points) and atomic inversion $J_z$ (blue points) versus state prep pulse duration.  Blue line is a fit to the blue points with the expected functional form of inversion versus state preparation angle, allowing for a fixed fraction of atoms that remain in the ground state.  Black line is predicted value for pulling coefficient given the fit parameters of the blue line, up to an overall scale factor.  Inset: the same data can be viewed by plotting $P$ against the value of $J_z'(0)$ inferred from the blue fit line, showing the expected linear scaling.  Figure reproduced from ref.~\cite{norcia2018cavity}. 
}
\label{fig:oat_exp_vsTheta}
\end{figure}

Next, we demonstrate the linear scaling of $P$ with effective atomic population inversion  $J_z^\prime$ by applying varying durations of preparation pulse and either measuring the resulting population inversion $J_z$ or toggling the cavity detuning $\delta_c$ by $\pm 2 \pi \times 30$~kHz and measuring the resulting change in the emission frequency fig.~\ref{fig:oat_exp_vsTheta}.  We observe the expected linear dependence of $P$ on $J_z'$.  

Because the cavity pulling switches signs at $J_z' = 0$, it is possible to create superradiant pulses with very low sensitivity to the cavity frequency by choosing an initial population inversion for which the time-averaged effect of the pulling nearly cancels.  This technique has enabled the observation of pulsed superradiance with a pulling coefficient of order $2 \times 10^{-6}$ \cite{norcia2018frequency}.  

This low sensitivity to cavity frequency shifts has enabled initial characterizations of the output frequency stability by comparison with a stable reference laser that is also compared to an optical lattice clock \cite{norcia2018frequency}.  The fractional frequency stability at one second of averaging was found to be $6.7(1)\times10^{-16}$, within roughly one order of magnitude of the state of the art for optical lattice clocks \cite{schioppo2016ultrastable, oelker2019optical}. After accounting for known systematic shifts, the absolute frequency of the emitted light was confirmed at the 2~Hz level to be in agreement with the expected atomic transition frequency.  

So far, all measurements performed on superradiant emission from an ultranarrow optical clock transition have used pulsed superradiance, where each atom contributes a single photon to the superradiant output.  After this superradiant pulse, the atoms are discarded, and a new ensemble prepared.  The performance of a superradiant frequency reference could be dramatically improved by operating in a steady-state manner.  This would mitigate limitations associated with Fourier broadening of the signal in the frequency domain, would improve limitations due to photon shot noise as more photons could be collected, and would eliminate noise aliasing associated with the experimental dead-time required to prepare a new atomic ensemble.  

Quasi steady-state operation has been achieved both in a Raman superradiant laser \cite{bohnet2012steady}, and in a laser based on the 7.5~kHz linewidth $^1$S$_0$ to $^3$P$_1$ transition in $^{88}$Sr \cite{norcia2015cold} by applying repump lasers to incoherently return atoms to the excited state and allow for multiple photons to be collected per atom.  Due to the abundance of metastable excited states present in fermionic alkaline earth atoms, such repumping schemes have yet to be demonstrated in a superradiant source operating on an ultranarrow linewidth clock transition.  

Even with repumping in place, true steady state operation would require a constant supply of atoms loaded into the cavity mode in order to compensate for atom loss and heating.  This in turn may require a device in which different steps of cooling and state preparation are separated in space, rather than in time as is currently standard in cold-atom experiments.  This challenge is currently being actively pursued both experimentally \cite{bennetts2017steady} and theoretically \cite{kazakov2014active}.

\section{Summary and Conclusion}
Narrow linewidth optical transitions in ions and atoms with two valence electrons have already lead to substantial advances in frequency metrology by providing exceedingly high quality atomic frequency references.  Combining these transitions with optical cavities provides new opportunities by enabling nondestructive atom counting techniques in optical lattice clocks, which may allow one to overcome limits associated with dead-time in spectroscopic sequences and quantum projection noise, by enhancing spectroscopy techniques for high-bandwidth laser frequency stabilization, and for developing new forms of active atomic frequency references.  

While proof of principle demonstrations of spin squeezing and nondestructive atom counting of atoms in cavities have been successfully performed using both narrow and broad linewidth optical transitions, \cite{Leroux10, braverman2019near, hosten2016measurement, cox2016deterministic}, it remains an open challenge to apply quantum enhancement techniques to a clock that operates in the optical domain.  One particular difficulty in doing so is to apply optical rotations with the agility and fidelity necessary to take advantage of entanglement, a task that is further complicated by the presence of a high-finesse cavity surrounding the atoms.  

For both cavity enhanced spectroscopy and active optical frequency references, a major outstanding goal is to move from an intermittent to continuous mode of operation \cite{bennetts2017steady}.  This becomes especially challenging for experiments involving ultra-narrow linewidth transitions, for which perturbations associated with atomic motion, light shifts and scattering from stray light must be controlled at a high level.  Such challenges provide exciting prospects for future experiments.

\noindent \textbf{Acknowledgements} 
I thank my thesis advisor James Thompson, experimental colleagues Matthew Winchester, Julia Cline, and Juan Muniz, and collaborators Ana Maria Rey, Robert Lewis-Swan, Bihui Zhu, Murray Holland, Jun Ye, John Robinson, Ross Hutson, Ed Marti, and Aki Goban for making the work presented here possible, and Dylan Young for a critical reading of the manuscript.

\end{document}